\documentclass[lettersize,journal]{IEEEtran}
\usepackage{orcidlink}
\usepackage{amsmath,amsfonts}
\usepackage{amssymb,physics}
\usepackage{algorithm}
\usepackage{array}
\usepackage[caption=false,font=normalsize,labelfont=sf,textfont=sf]{subfig}
\usepackage{textcomp}
\usepackage{stfloats}
\usepackage{url}
\usepackage{verbatim}
\usepackage{graphicx}
\usepackage{cite}
\usepackage[nolist]{acronym}

\usepackage{color}
\usepackage{float}
\usepackage{setspace}
\usepackage[noend]{algpseudocode}

\usepackage{caption}
\usepackage{subcaption}
\usepackage{xcolor}

\usepackage{comment}
\usepackage{cite}
\usepackage{soul}
\usepackage{epstopdf}
\usepackage{xfrac}
\usepackage{pifont}
\newcommand{\cmark}{\textcolor{green}{\ding{51}}}%
\newcommand{\xmark}{\textcolor{red}{\ding{55}}}%

\usepackage{csquotes}
\usepackage{booktabs}
\usepackage{multirow}
\usepackage{tabularx,ragged2e}
\newcolumntype{Y}{>{\centering\arraybackslash}X}
\usepackage{footmisc}

\usepackage{varwidth}
\usepackage{tikz}
\usepackage{pgf}
\usepackage{pgfplots}
\usepackage{pgfplotstable}
\usepackage{hyperref}
\usepackage{cleveref}

\usepackage{svg}
\usepackage{varwidth}
\usepackage{tikz}
\usepackage{pgf}
\usepackage{pgfplots}
\usepackage{pgfplotstable}
\usetikzlibrary{positioning,backgrounds,fit,calc,patterns,arrows.meta}
\pgfplotsset{compat=1.3}
\usetikzlibrary{shapes.geometric}
\usetikzlibrary{shapes.arrows}
\usetikzlibrary{arrows.meta}
\usepackage[skins]{tcolorbox}
\usepackage{adjustbox}

\pgfplotsset{compat=1.3}
\usetikzlibrary{shapes.geometric}
\usetikzlibrary{shapes.arrows}
\usepackage{array}

\newcommand*\diff{\mathop{}\!\mathrm{d}}
\newcommand{\D}[0]{\mathrm{d}}
\newcommand{\state}[0]{\mathbf{x}_\tau}
\newcommand{\wien}[0]{\mathbf{w}_\tau}
\newcommand{\clean}[0]{\mathbf{x}_0}
\newcommand{\init}[0]{\mathbf{x}_T}
\newcommand{\reverb}[0]{\mathbf{y}}
\newcommand{\sco}[0]{\nabla_{\state} \log p(\state)}

\newcommand{\rir}[0]{\mathbf{h}}

\newcommand{\postsconopsi}[0]{\nabla_{\state} \log p(\state | \reverb)}

\newcommand{\likelihoodnopsi}[0]{\nabla_{\state} \log p(\reverb | \state)}
\newcommand{\scomodel}[0]{\mathbf{s}_\theta(\state, \tau)}
\newcommand{\scomodelopti}[0]{\mathbf{s}_{\theta^*}(\state, \tau)}
\newcommand{\kernel}[0]{q_\tau( \state | \clean)}
\newcommand{\tweedie}[0]{\hat{\mathbf{x}}_0^\tau}

\begin{acronym}
\acro{stft}[STFT]{short-time Fourier transform}
\acro{istft}[iSTFT]{inverse short-time Fourier transform}
\acro{dnn}[DNN]{deep neural network}
\acro{pesq}[PESQ]{Perceptual Evaluation of Speech Quality}
\acro{polqa}[POLQA]{perceptual objectve listening quality analysis}
\acro{wpe}[WPE]{weighted prediction error}
\acro{psd}[PSD]{power spectral density}
\acro{rir}[RIR]{room impulse response}
\acro{snr}[SNR]{signal-to-noise ratio}
\acro{lstm}[LSTM]{long short-term memory}
\acro{polqa}[POLQA]{Perceptual Objective Listening Quality Analysis}
\acro{sdr}[SDR]{signal-to-distortion ratio}
\acro{estoi}[ESTOI]{extended short-term objective intelligibility}
\acro{elr}[ELR]{early-to-late reverberation ratio}
\acro{tcn}[TCN]{temporal convolutional network}
\acro{rls}[RLS]{recursive least squares}
\acro{asr}[ASR]{automatic speech recognition}
\acro{ha}[HA]{hearing aid}
\acro{ci}[CI]{cochlear implant}
\acro{mac}[MAC]{multiply-and-accumulate}
\acro{gan}[GAN]{generative adversarial network}
\acro{tf}[T-F]{time-frequency}
\acro{sde}[SDE]{stochastic differential equation}
\acro{ode}[ODE]{ordinary differential equation}
\acro{drr}[DRR]{direct to reverberant ratio}
\acro{lsd}[LSD]{log spectral distance}
\acro{sisdr}[SI-SDR]{scale-invariant signal to distortion ratio}
\acro{mos}[MOS]{mean opinion score}
\acro{map}[MAP]{maximum a posteriori}
\acro{sde}[SDE]{stochastic differential equation}
\acro{ode}[ODE]{ordinary differential equation}
\acro{dps}[DPS]{diffusion posterior sampling}
\acro{vae}[VAE]{variational auto-encoding}
\acro{nmf}[NMF]{non-negative matrix factorization}
\acro{wpe}[WPE]{weighted-prediction error}
\acro{em}[EM]{expectation-maximization}
\end{acronym}

\definecolor{statedps}{HTML}{FFC300}
\definecolor{dps}{HTML}{EC610A}
\definecolor{kodrasi}{HTML}{000000}

\makeatletter
\newcommand{\subalign}[1]{%
  \vcenter{%
    \Let@ \restore@math@cr \default@tag
    \baselineskip\fontdimen10 \scriptfont\tw@
    \advance\baselineskip\fontdimen12 \scriptfont\tw@
    \lineskip\thr@@\fontdimen8 \scriptfont\thr@@
    \lineskiplimit\lineskip
    \ialign{\hfil$\m@th\scriptstyle##$&$\m@th\scriptstyle{}##$\hfil\crcr
      #1\crcr
    }%
  }%
}
\makeatother

\begin{document}

\title{Unsupervised Blind Joint Dereverberation and Room Acoustics Estimation with Diffusion Models}

\author{
Jean-Marie Lemercier$^{\ast}$\,{\orcidlink{0000-0002-8704-7658}}, \IEEEmembership{Student Member,~IEEE}, 
Eloi Moliner$^{\ast}$\,{\orcidlink{0000-0001-5719-326X}},
 Simon Welker\,{\orcidlink{0000-0002-6349-8462}},
 ~\IEEEmembership{Student Member,~IEEE}, 
Vesa Välimäki\,{\orcidlink{0000-0002-7869-292X}},
\IEEEmembership{Fellow,~IEEE},
Timo Gerkmann\,{\orcidlink{0000-0002-8678-4699}},
\IEEEmembership{Senior Member,~IEEE}
\thanks{
$^{\ast}$Equal contribution. 
Jean-Marie Lemercier, Simon Welker and Timo Gerkmann are with the Signal Processing group at Universit\"at Hamburg, Hamburg, Germany. 
Eloi Moliner and Vesa Välimäki are with the Acoustics Lab, Department of Information Communications Engineering, Aalto University, Espoo, Finland.
The authors gratefully acknowledge the computing resources provided by both the Erlangen National High Performance Computing Center (NHR@FAU) of the Friedrich-Alexander-Universität Erlangen-Nürnberg (FAU) (NHR project F101AC1) and the Aalto Science-IT project.
}
}

\maketitle

\begin{abstract}
This paper presents an unsupervised method for single-channel blind dereverberation and room impulse response (RIR) estimation, called BUDDy. The algorithm is rooted in Bayesian posterior sampling: it combines a likelihood model enforcing fidelity to the reverberant measurement, and an anechoic speech prior implemented by an unconditional diffusion model.
We design a parametric filter representing the RIR, with exponential decay for each frequency subband.
Room acoustics estimation and speech dereverberation are jointly carried out, as the filter parameters are iteratively estimated and the speech utterance refined along the reverse diffusion trajectory.
In a blind scenario where the RIR is unknown, BUDDy successfully performs speech dereverberation in various acoustic scenarios, significantly outperforming other blind unsupervised baselines.
Unlike supervised methods, which often struggle to generalize, BUDDy seamlessly adapts to different acoustic conditions.
This paper extends our previous work by offering 
new experimental results and insights into the algorithm’s versatility. 
We demonstrate the robustness of our proposed method to new acoustic and speaker conditions, as well as its adaptability to high-resolution singing voice dereverberation, using both instrumental metrics and subjective listening evaluation.
We study BUDDy's performance for RIR estimation and observe it surpasses a state-of-the-art supervised DNN-based estimator on mismatched acoustic conditions.
Finally, we investigate the sensitivity of informed dereverberation methods to RIR estimation errors, thereby motivating the joint acoustic estimation and dereverberation design.
Audio examples and code can be found online.\footnote{\href{uhh.de/sp-inf-buddy}{uhh.de/sp-inf-buddy}}

\end{abstract}

\begin{IEEEkeywords}
Acoustics, reverberation, speech enhancement.
\end{IEEEkeywords}

\section{Introduction}

\IEEEPARstart{R}{everberation} is a natural phenomenon caused by acoustic waves propagating in a space and reflecting off various surfaces, such as walls, ceilings, floors, and objects within the environment.
Reverberation and particularly late reflections often degrade speech intelligibility and quality for normal listeners, and even more severely so for hearing-impaired listeners \cite{Naylor2011}. 
Therefore, many communication devices now include a dereverberation algorithm, which aims to recover the anechoic component of speech. 
This paper considers the case in which recordings from only one microphone are available, which is more challenging than a multi-channel case \cite{Miyoshi1988MINT}.

Traditional dereverberation algorithms operate in the time, spectral, or cepstral domain~\cite{gerkmann2018book_chapter}, leveraging statistical assumptions about the anechoic and reverberant signals \cite{gerkmann2010} as well as properties of the reverberation signal model \cite{Kodrasi2014}.
Two scenarios are considered for dereverberation, depending on the knowledge of the room acoustics represented by the \ac{rir}. 
Some methods tackle \textit{informed} scenarios, where the \ac{rir} is known \cite{Mourjopoulos1982Comparative, Kodrasi2014}, whereas other approaches consider \textit{blind} scenarios where the \ac{rir} is unknown 
\cite{Nakatani2008a, yohena2024single, Schmid2012AMaximum, Jukic2014speech, Habets2010}.
Informed dereverberation is naturally an easier task than blind dereverberation. However, knowing the \ac{rir} does not guarantee obtaining a stable and causal inverse filter in the single-channel case, since real-world RIRs are mixed-phase systems \cite{Neely1979Invertibility}. 
Using multiple microphones helps resolve this issue to some extent \cite{Miyoshi1988MINT}, but informed dereverberation methods generally exhibit other weaknesses such as a lack of robustness to RIR estimation errors \cite{Hikichi2007Inverse}.
Additionally, most scenarios in real-life applications are (at least partially) blind, as the \ac{rir} is either not measured beforehand, or only valid for a specific acoustic setting.

Data-driven approaches rely less on distributional assumptions than statistical methods but instead directly learn the signal properties and structures from data \cite{wang2018supervised}. 
Most of these methods are based on supervised learning, where models are trained using paired data. Each input is associated with a corresponding target output, allowing the model to learn a mapping from inputs to outputs.
For dereverberation, this typically involves using pairs of anechoic and reverberant speech, where the latter is often produced by convolving anechoic speech signals with \ac{rir}s.
Supervised predictive models are particularly popular for blind dereverberation: these range from time-frequency masking \cite{Williamson2017j} and mapping\cite{Han2017} to algorithms operating on the cepstrum \cite{liu2024dual} or directly on the waveform \cite{Ernst2019, zhao2020}.

Generative modeling is another paradigm gaining a lot of interest in audio restoration tasks \cite{lemercier2024restoration}, including dereverberation.
Generative models for speech dereverberation learn a parameterization of the posterior distribution of clean speech conditioned on reverberant speech.
Diffusion models in particular \cite{sohl2015deep, ho2020denoising, song2021sde} have been extensively investigated for such conditional generation task,
leading to the introduction of diffusion-based blind supervised dereverberation algorithms \cite{richter2023speech, Lemercier2022storm}. 
Still, the generalization ability of supervised approaches is limited by their design.

In contrast, unsupervised methods operate without paired data, relying solely on patterns learned from anechoic speech signals.
These approaches have been getting little visibility but boast interesting properties such as improved robustness to unseen acoustic conditions without the need for retraining. 
An unsupervised method for informed single-channel dereverberation based on diffusion models was proposed in our prior work \cite{lemercier2023derevdps}. 
That approach is based on Bayesian diffusion posterior sampling (DPS) \cite{chung_diffusion_2022}, combining a diffusion-based anechoic speech prior and a Gaussian likelihood model for state-of-the-art informed dereverberation.
However, as shown in this work, such an informed algorithm is sensitive to even small \ac{rir} estimation errors, rendering it impractical in real-life scenarios. 

Related works in other signal processing domains have already considered blind inverse problems through the lens of posterior sampling with diffusion priors.
For image deblurring, Chung et al.~\cite{chung_parallel_2022} propose to use an additional diffusion process dedicated to estimating the deblurring kernel, while Laroche et al.~\cite{laroche2023fast} adapts an expectation-maximization algorithm using a denoising regularization of the blurring kernel, and Sanghvi et al.~\cite{sanghvi2023kernel} dedicates a non-blind solver to estimate a deblurred image at each diffusion step.
For speech denoising, Nortier et al.~\cite{nortier2023unsupervised} combine a noise model based on non-negative matrix factorization with a clean speech diffusion prior.
Moliner et al.~\cite{moliner2024blind} address the problem of blind bandwidth extension by leveraging a diffusion prior and iteratively optimizing a parametric lowpass filter operator.
Recent works adapt denoising diffusion restoration models (DDRM) \cite{kawar2022ddrm} for singing voice dereverberation \cite{saito2023unsupervised, murata_gibbsddrm_2023}, using an initialization provided by the \ac{wpe} algorithm \cite{Nakatani2008a}.

For speech dereverberation, a first generative model based on traditional Gaussian mixtures was proposed in \cite{Attias2000speech}. Other works learn an anechoic speech prior via \ac{vae}: the VAE-NMF method \cite{Baby2021vaenmf} models reverberation via non-negative matrix factorization and estimates its parameters with a Monte-Carlo method; the RVAE-EM model \cite{Wanf2024rvaeem} adopts a 
maximum a posteriori
perspective, combining a recurrent VAE prior with a Gaussian likelihood model.
Unsupervised dereverberation with a non-generative prior has also been investigated in the multi-channel scenario \cite{wang2024usdnetunsupervisedspeechdereverberation}.

This paper expands our prior work \cite{moliner2024buddy}, where we designed a blind unsupervised dereverberation algorithm, extending \cite{lemercier2023derevdps} to the blind scenario. The resulting approach, called BUDDy, uses a model-based parametric subband filter with an exponential decay to approximate the \ac{rir}. 
BUDDy performs joint estimation of the \ac{rir} and the anechoic speech, leveraging the model-based parameterization as an acoustic prior and the diffusion model as a speech prior. 
We have shown previously \cite{moliner2024buddy} that BUDDy can successfully remove reverberation, and that it is robust to changes in acoustic conditions because of the lack of supervision during training. 
Therefore, BUDDy closes the performance gap between matched and mismatched acoustic conditions in comparison to diffusion-based supervised approaches \cite{richter2023speech, Lemercier2022storm}.

In this paper, we extend the experimental framework of our previous publication \cite{moliner2024buddy} with the following contributions:
\begin{itemize}
\item  Section \ref{sec:vctk} extends the evaluation of BUDDy for speech dereverberation beyond instrumental metrics, including a subjective listening test and a set of ablation studies. Notably, we perform experiments on speech recordings made in real environments, rather than using the convolution model, a scenario in which BUDDy performs especially well.
\item  Section \ref{sec:nhss} presents new experiments on applying BUDDy to singing voice dereverberation at a sampling rate of 44.1 kHz, which is higher than the 16-kHz sampling rate used in our speech experiments \cite{moliner2024buddy}. The results, which also include a subjective listening test, indicate that our method significantly outperforms existing unsupervised state-of-the-art approaches and performs comparably to supervised baselines.
\item In Section \ref{sec:robustness}, we investigate the \emph{robustness} of informed dereverberation approaches in partially blind scenarios, in comparison to BUDDy. We highlight the limitations of these approaches when the RIR is perturbed with Gaussian noise or estimated blindly using a state-of-the-art RIR estimator \cite{steinmetz2021filtered}.
\item  Finally, Section \ref{sec:rir} assesses BUDDy’s performance in \emph{RIR estimation} against a state-of-the-art supervised estimator \cite{steinmetz2021filtered}. We use frequency-wise acoustic descriptors to evaluate the accuracy of BUDDy on reverberation time and clarity.
\end{itemize}

We organize the paper as follows. 
In Section \ref{sec:background}, we introduce diffusion-based generative models and posterior sampling methods for informed dereverberation using diffusion priors as proposed in previous work \cite{lemercier2023derevdps}. Then in Section \ref{sec:method}, we introduce our blind unsupervised dereverberation method BUDDy \cite{moliner2024buddy}, which extends the posterior sampling method presented in the previous section to the blind scenario where the \ac{rir} is not available.
The experiments and results mentioned above are presented in Section \ref{sec:results}.
Section \ref{sec:conclusion} concludes the paper.

\label{sec:introduction}

 \section{Informed Diffusion-Based Dereverberation} \label{sec:background}

 This section introduces diffusion models, a class of generative models that form the foundation of the proposed method. It also explores their application in solving inverse problems, specifically highlighting their use in informed dereverberation.
 
 Throughout this paper, we use the following notations: considering dereverberation under the prism of inverse problem solving, we wish to retrieve the anechoic time-domain utterance $\clean \in \mathbb{R}^L$, where $L$ is the length of the utterance, given the reverberant measurement $\reverb$. As in most dereverberation studies, reverberation is modeled as a convolution between anechoic speech with a \ac{rir} $\rir \in \mathbb{R}^{L_\mathbf{h}}$, such that $\reverb = \rir \ast \clean$, where $\ast$ is the discrete convolution operator in the time domain, resulting in $\reverb \in \mathbb{R}^{L+L_\mathbf{h}-1}$.

 \subsection{Diffusion-Based Generative Models}

\label{sec:basics_sgm}
Diffusion models \cite{ho2020denoising, song2019generative} have 
achieved remarkable success across various domains, including speech \cite{kong2021diffwave}. 
They break down the problem of generating high-dimensional complex data into a series of easier denoising tasks. 
Training a diffusion model first requires defining a \emph{forward process}, which gradually adds noise to data points, turning the target data distribution into a tractable Gaussian distribution.

The forward process is the solution of the following \ac{sde}:
\begin{equation}\label{eq:forward-sde}
    \D \state = \mathbf{f}( \state,\tau) \D \tau + g(\tau) \D \mathbf{w}_\tau \,, 
\end{equation}
where the diffusion time $\tau$ indexes the stochastic process $\mathbf{x}_\tau$ from $\tau=T_\mathrm{min}$ to $\tau= T \gg T_\mathrm{min}$. 
The minimal process time $T_\mathrm{min}$ is chosen stricly positive to avoid irregularities around $0$.
The Wiener process $\mathbf{w}_\tau$ injects noise with independent and normally distributed increments, that is,
$\mathbf{w}_{\tau + \diff{\tau}} - \wien 
\sim \mathcal{N}(\mathbf{0}, \diff{\tau} \, \mathbf{I})$ where $\mathbf{I} \in \mathbb{R}^{L \times L}$ is the identity matrix \cite{Oksendal2000SDE}.
The diffusion state $\state \in \mathbb{R}^L$ starts at a clean speech data point $\clean \in \mathbb{R}^L \sim p_\text{data}$ and ends at the final state $\init \in \mathbb{R}^L$ which contains mostly Gaussian noise.
We adopt the parameterization proposed by Karras et al. \cite{karras2022elucidating}, which defines the \emph{drift} and \emph{diffusion} parameters as $f(\mathbf{\state}, \tau) = 0$ and $g(\tau) = \sqrt{2 \tau}$, respectively. 
This results in a noise schedule $\sigma(\tau)=\tau$ which determines the so-called \textit{transition kernel} i.e. the marginal density of the forward process \cite{karras2022elucidating}
\begin{equation} \label{eq:transition}
    q_\tau(\state | \clean) = \mathcal{N}(\clean, \sigma^2(\tau) \mathbf{I}) \,.
\end{equation}

Conversely, data generation is accomplished by reversing the forward corruption process. First, an initial sample is drawn from a Gaussian distribution, and then the model iteratively removes noise until a clean sample from the target distribution emerges.
The \textit{reverse process}, 
can be characterized by the \emph{probability flow} \ac{ode}, which has the same marginal distributions as the reverse \ac{sde} canonically associated to the forward \ac{sde} \eqref{eq:forward-sde} \cite{anderson1982reverse}
\begin{equation}\label{eq:ode}
    \D \state = \left[ \mathbf{f}( \state,\tau)  - 
     \frac{g^2(\tau)}{2}
     \sco \right] \D \tau \,, 
\end{equation}
where diffusion time $\tau$ flows in reverse from $\tau=T$ to $\tau=T_\mathrm{min}$. 
The diffusion state $\state \in \mathbb{R}^L$ starts from the initial state $\init \in \mathbb{R}^L$
and ends at $\clean \in \mathbb{R}^L \sim p_\text{data}$.
The \emph{score function} $\sco$ indicates the direction towards regions of higher probability under the model’s distribution. 
In practice, it is intractable and we need to estimate it with a \textit{score model} $\scomodel$ parameterized with a \ac{dnn}.
Vincent et al. have shown that the score model $\scomodel$ can be optimized using denoising score matching \cite{vincent2011connection}, i.e. matching the score of the Gaussian transition 
kernel $\kernel$ instead of the score of the unknown probability $p(\state)$. 
The score of the transition kernel $\kernel$ can be obtained from \eqref{eq:transition} as
\begin{equation} \label{eq:score}
    \nabla_{\state} \log \kernel = - \frac{\state - \clean }{\sigma^2(\tau)} \,.
\end{equation}
The score model $\mathbf{s}_\theta$ is therefore trained using the denoising score-matching objective \cite{vincent2011connection}
\begin{equation} \label{eq:training}
\mathbb{E}_{\subalign{&\tau \sim \mathcal{U}(T_\text{min}, T_\text{max}) \\ &\clean \sim p_\text{data} \\ &\state \sim \kernel}}
\left[
 \lambda(\tau) 
 \left\lVert
 \scomodel + \frac{\state - \clean }{\sigma^2(\tau)} 
 \right\rVert_2^2
 \right],
\end{equation}
where first a diffusion index $\tau$ is randomly sampled 
between extremal times $T_\mathrm{min}$ and $T_\mathrm{max} > T$, a data point $\clean$ is sampled in the training set, and the corresponding diffusion state $\state$ is obtained from the transition kernel in \eqref{eq:transition}. 
In practice, we use the same pre-conditioning for $\scomodel$ and same loss weighting $\lambda(\cdot)$ as in Karras et al. (see \cite{karras2022elucidating} for details).

 \subsection{Diffusion Posterior Sampling for Dereverberation} \label{sec:dps}
 
We discuss in this section how diffusion priors can be adapted in order to solve inverse problems. 
While some traditional methods derive maximum a posteriori estimators for blind dereverberation
\cite{Schmid2012AMaximum, Jukic2014speech, Habets2010}, 
we exploit the generative nature of diffusion models to solve this inverse problem using posterior sampling.
Assuming that the RIR $\mathbf{h}$ is known, we attempt to sample from the posterior distribution of the anechoic speech given the measurement and the RIR $p(\clean | \reverb) \propto p(\reverb | \clean) p(\clean)$.
Although we do not have an explicit prior $p(\clean)$ like in e.g. \ac{vae} frameworks, we leverage the implicit prior $p_{\theta^*}(\clean)$ given by the pretrained diffusion model $\scomodelopti$ where $\theta^*$ represents the (fixed) parameters optimizing the training objective \eqref{eq:training}.
Sampling is then achieved by solving the 
probability flow ODE \eqref{eq:ode}, 
replacing the unconditional score function by the \emph{posterior score} $\postsconopsi$ \cite{song2021sde} obtained through Bayes' rule
\begin{align} \label{eq:reverse-diff-estep}
    \D \state &= 
    \left[ \mathbf{f}( \state,\tau) - \frac{1}{2} g^2(\tau) \postsconopsi \right] \D \tau \,, \nonumber \\
 & \mspace{-13mu}
 \approx \left[ \mathbf{f} \left( \state,\tau \right) - \frac{1}{2} g^2(\tau) \left( \scomodelopti + \likelihoodnopsi \right) \right] \D \tau \,.
\end{align}
The \textit{likelihood score} $\likelihoodnopsi$ is in general intractable for $\tau>0$. Following \cite{chung_diffusion_2022}, we employ a plug-in estimate of $\clean$ denoted as $\tweedie$ which we derive using Tweedie's formula, i.e. one-step denoising of $\state$ using the diffusion model
\begin{equation} \label{eq:tweedie}
    \tweedie 
    \overset{\Delta}{=}
    \mathbb{E}[\clean | \state] \approx
    \state + \sigma^2(\tau) \scomodelopti \,.
\end{equation}
We assume that this estimate is a sufficient statistic for $\state$, which results in a first assumption $p(\reverb | \state) \approx p(\reverb | \tweedie)$.

In order to approximate $p(\reverb | \tweedie)$, previous work \cite{lemercier2023derevdps} models the error between $\reverb$ and its estimation
to follow a zero-mean Gaussian distribution in the time domain. 
The corresponding expression for the likelihood score $\likelihoodnopsi$ is then 
a simple weighted $L^2$-distance between $\reverb$ and $\rir \ast \tweedie$.
However 
far better dereverberation performance and speech quality can be achieved by substituting the obtained distance with a $L^2$-distance between compressed \ac{stft} representations instead.
This is analogous to modeling the likelihood as
\begin{equation}\label{eq:likelihood}
    p(\reverb | \clean) \propto \mathrm{exp} \big( - \zeta(\tau) 
\, \mathcal{C}(\reverb, \rir \ast \tweedie )\, \big) \,,
\end{equation}
where the scalar $\zeta(\tau)$ 
controls the influence of the likelihood score term during sampling,
and $\mathcal{C}(\cdot, \cdot)$ is the cost function
\begin{equation}\label{eq:objective}
\mathcal{C}(\mathbf{u}, \mathbf{v}) = \frac{1}{M}
\sum_{m=1}^M
\sum_{k=1}^K 
\rVert
 S_\mathrm{comp}(
\mathbf{u})
_{m,k}-
 S_\mathrm{comp}(
\mathbf{v}
)
_{m,k}
\lVert_2^2 \,.
\end{equation}
There, 
 $S_\mathrm{comp}(
\mathbf{u}) \in \mathbb{C}^{M \times K}$ denotes the magnitude-compressed \ac{stft} of $\mathbf{u}$,
comprising $M$ time frames and $K$ frequency bins
\begin{equation}\label{eq:spec_comp}
     S_\mathrm{comp}(
    \mathbf{u})=|\text{STFT}(\mathbf{u})|^{2/3} \exp{j \angle \text{STFT}(\mathbf{u})} \,.
\end{equation}
We apply this compression to boost low-energy components as typically observed in high frequencies of speech signals or in late reverberation tails, and account for the heavy-tailedness of speech distributions \cite{gerkmann2010}.
Such a strategy is also employed in \cite{Welker2022SGMSE} for data representation.
In our case, we only use this non-linear transformation in the cost function \eqref{eq:objective}, whereas the diffusion process itself uses the original time-domain representation.

The parameter $\zeta(\tau)$ balances a trade-off between adherence to the prior data distribution and fidelity to the observed data.
We empirically resort to the same parameterization of $\zeta(\tau)$ as in \cite{moliner_solving_2022, moliner2024blind}:
\begin{equation} \label{eq:variance}
    \zeta(\tau) = \frac{\sqrt{L} \, \tilde{\zeta}}{ 
    \sigma(\tau)
    \norm{\nabla_{\state} \mathcal{C} \left( \mathbf{y}, \rir \ast \hat{\mathbf{x}}_0(\state) \right) }_2 }  \, ,
\end{equation}
where $\tilde{\zeta}$ is a fixed coefficient.

The resulting informed dereverberation algorithm is a slight variation of our previous work \cite{lemercier2023derevdps}. In the following we refer to this approach as InfDerevDPS.

\section{Blind Diffusion-Based Dereverberation}\label{sec:method}

\begin{algorithm}[t]
\caption{Reverberation Operator $\mathcal{A}_\psi (\cdot)$}
\label{alg:operator}
\begin{algorithmic}
\Function{$\mathcal{A}_\psi$}{$\hat{\mathbf{x}}_0$}
    \State $\{\mathbf{\Phi}, (w_b, \alpha_b)_{b=1,\dots, B} \} \gets \psi$ \Comment{Parameter set}
    \State $\mathbf{A}_{n,b}^\prime \gets w_b \cdot e^{-\alpha_b n }$ \Comment{Exponential decay model}
    \State $\mathbf{A} \gets \exp{\left( \mathrm{lerp}( \log{\mathbf{A}^\prime}) \right)}$ 
    \Comment{Frequency interpolation}
    \State $\mathbf{H} \gets \mathbf{A} \cdot e^{j \mathbf{\Phi}}$
    \State $\overline{\mathbf{H}} \gets \mathrm{STFT}(\delta_\text{d} \oplus \mathcal{P}_\text{min}(\mathrm{iSTFT}(\mathbf{H})))$  \Comment{Projection step}
    \State $\hat{\mathbf{X}} \gets \mathrm{STFT}(\hat{\mathbf{x}}_0)$
    \State $\hat{\mathbf{Y}}_{m,k} \gets \sum_{n=0}^{N_h} \overline{\mathbf{H}}_{n,k} \hat{\mathbf{X}}_{m-n,k}$ \Comment{Subband convolution}
    \State \Return $\mathrm{iSTFT}(\hat{\mathbf{Y}})$
\EndFunction
\end{algorithmic}
\end{algorithm}

\definecolor{cb1}{HTML}{D81B60}
\definecolor{cb2}{HTML}{1E88E5}
\definecolor{cb3}{HTML}{D29E02}
\definecolor{cb4}{HTML}{004D40}
\definecolor{cb5}{HTML}{864dbf}

\newcounter{phase}[algorithm]
\newlength{\phaserulewidth}
\newcommand{\setphaserulewidth}{\setlength{\phaserulewidth}}
\newcommand{\phase}[1]{%
  \vspace{0.1em}
  \Statex\strut\refstepcounter{phase}
  \hspace{1em}\textit{Phase~\thephase~--~#1}%
  \vspace{-2ex}
  \Statex\leavevmode\llap{ 
  }
  \hspace{1em} 
  \rule{0.9\linewidth}{\phaserulewidth}
}
\makeatother
\setphaserulewidth{.3pt}

\begin{algorithm}[t]
\caption{Inference algorithm}
\label{alg:inference}
\begin{algorithmic}
\Require Reverberant speech $\mathbf{y}$
\State $\mathbf{x}_\mathrm{init} \leftarrow \mathrm{\textcolor{cb5}{WPE}}(\mathbf{y})$  
\State Sample $\mathbf{x}_{N}\sim \mathcal{N}({\mathbf{x}_\mathrm{init}},\sigma_N^2\mathbf{I})$ \Comment{Warm initialization}
\State Initialize $\psi_{N}$ \Comment{Initialize the RIR parameters}
\For{$n \leftarrow N, \dots, 1$} \Comment{Discrete step backwards}

\phase{E-step}

\State $\mathbf{s}_n \leftarrow s_\theta(\mathbf{x}_n, \tau_n) $ \Comment{\textcolor{cb1}{Evaluate score model}}
\State $\hat{\mathbf{x}}_0 \leftarrow  \mathbf{x}_n - \sigma_n^2 \mathbf{s}_n$ \Comment{Get one-step denoising estimate}
\State $\hat{\mathbf{x}}_0 \leftarrow \mathrm{Rescale}(\hat{\mathbf{x}}_0) $ \Comment{Constraint RMS power}

\State $ \mathbf{g}_n \leftarrow - \zeta(\tau_n) \nabla_{\mathbf{x}_n} 
    \mathcal{C}(\reverb,
    \mathcal{A}_{\psi_{n}}(\hat{\mathbf{x}}_0
    ))$ \Comment{\textcolor{cb4}{LH score approx.}}
\State $\mathbf{x}_{{n-1}} \leftarrow \mathbf{x}_{n} - \sigma_n (\sigma_{n-1}-\sigma_n) (\mathbf{s}_n +\mathbf{g}_n)$ \Comment{\textcolor{cb3}{Update step}}

\phase{M-step}

\State $\psi_{n-1}^0 \leftarrow \psi_{n}$  \Comment{Use RIR parameters from last step}
\For{$j \leftarrow 0, \dots, N_\text{its.}$} \Comment{\textcolor{cb2}{RIR optimization}}
    \State $\mathcal{J}_\text{RIR}(\psi_{n-1}^j) \leftarrow \mathcal{C}(\mathbf{y},\mathcal{A}_{\psi_{n-1}^j}(\hat{\mathbf{x}}_0)) + \mathcal{R}(\psi_{n-1}^j)$
\State $\psi_{n-1}^{j+1} \leftarrow \psi_{n-1}^{j} - \mathrm{Adam}(\mathcal{J}_\text{RIR}(\psi_{n-1}^{j}))$  \Comment{Opti. step}
 \State $\psi_{n-1}^{j+1} \leftarrow \mathrm{clamp}(\psi_{n-1}^{j+1})$ \Comment{Constrain Parameters}
\EndFor
\State $\psi_{n-1} \leftarrow \psi_{n-1}^M$

\EndFor
\State \Return $\mathbf{x}_0$  \Comment{Reconstructed audio signal}
\end{algorithmic}
\end{algorithm}

\definecolor{cb1}{HTML}{D81B60}
\definecolor{cb2}{HTML}{1E88E5}
\definecolor{cb3}{HTML}{D29E02}
\definecolor{cb4}{HTML}{004D40}
\definecolor{cb5}{HTML}{864dbf}

\tikzstyle{mycircle} = [circle, draw, fill=white, inner sep=0pt, minimum size=25pt]
\tikzstyle{mysquare} = [rectangle, draw, fill=black, inner sep=0.1pt, minimum width=35pt, minimum height=35pt, align=center]
\tikzstyle{mybranch} = [circle, draw, fill=black, inner sep=0pt, , minimum size=2pt]
\tikzstyle{myrectangle} = [rectangle, draw, fill=cb1!20, inner sep=3pt, minimum width=35pt, minimum height=20pt, align=center]
\tikzstyle{myrectangle2} = [rectangle, draw, fill=cb2!20, inner sep=3pt, minimum width=50pt, minimum height=20pt, align=center]
\tikzstyle{myrectangle4} = [rectangle, draw, fill=cb3!20, inner sep=3pt, minimum width=50pt, minimum height=30pt, align=center]
\tikzstyle{myrectangle3} = [rectangle, draw, fill=cb4!20, inner sep=3pt, minimum width=30pt, minimum height=25pt, align=center]
\tikzstyle{myrectangle5} = [rectangle, draw, fill=cb5!20, inner sep=3pt, minimum width=30pt, minimum height=25pt, align=center]

\tikzstyle{myrectangle_white} = [rectangle, draw, fill=white, inner sep=3pt, minimum width=10em, minimum height=20pt, align=center]

\tikzstyle{sum} = [
  circle,
  draw,
  minimum size=9pt,
  append after command={
    \pgfextra{
      \draw (\tikzlastnode.north) -- (\tikzlastnode.south);
      \draw (\tikzlastnode.west) -- (\tikzlastnode.east);
    }
  },
]

\tikzstyle{product} = [
  circle,
  draw,
  minimum size=12pt,
  append after command={
    \pgfextra{
      \draw (\tikzlastnode.north east) -- (\tikzlastnode.south west);
      \draw (\tikzlastnode.north west) -- (\tikzlastnode.south east);
    }
  },
]

\tikzstyle{convolutioncircle} = [
  circle,
  draw,
  minimum size=24pt,
]

\newcommand{\specwidth}{1.35cm}
\newcommand{\sepnodex}{0.5cm}
\newcommand{\sepnodey}{0.4cm}
\newcommand{\sepnodexspec}{0cm}
\newcommand{\sepnodeyspec}{0.25cm}
\newcommand{\sepstate}{0.75cm}
\definecolor{figblue}{HTML}{154c79}
\definecolor{figgreen}{HTML}{a8db33}
\definecolor{figgrey}{HTML}{797979}

\begin{figure*}
\centering
\begin{tikzpicture}[scale=0.89, transform shape]
    
    \coordinate (stateN) at (0,0.0);

    \node[mycircle] (sN) at (stateN) {$\mathbf{x}_{N}$};
    \node[mycircle, below=3.6cm of sN] (spsiN) {$\psi_{N}$};
    \node[mysquare, below=\sepnodeyspec of sN, anchor=north, fill overzoom image={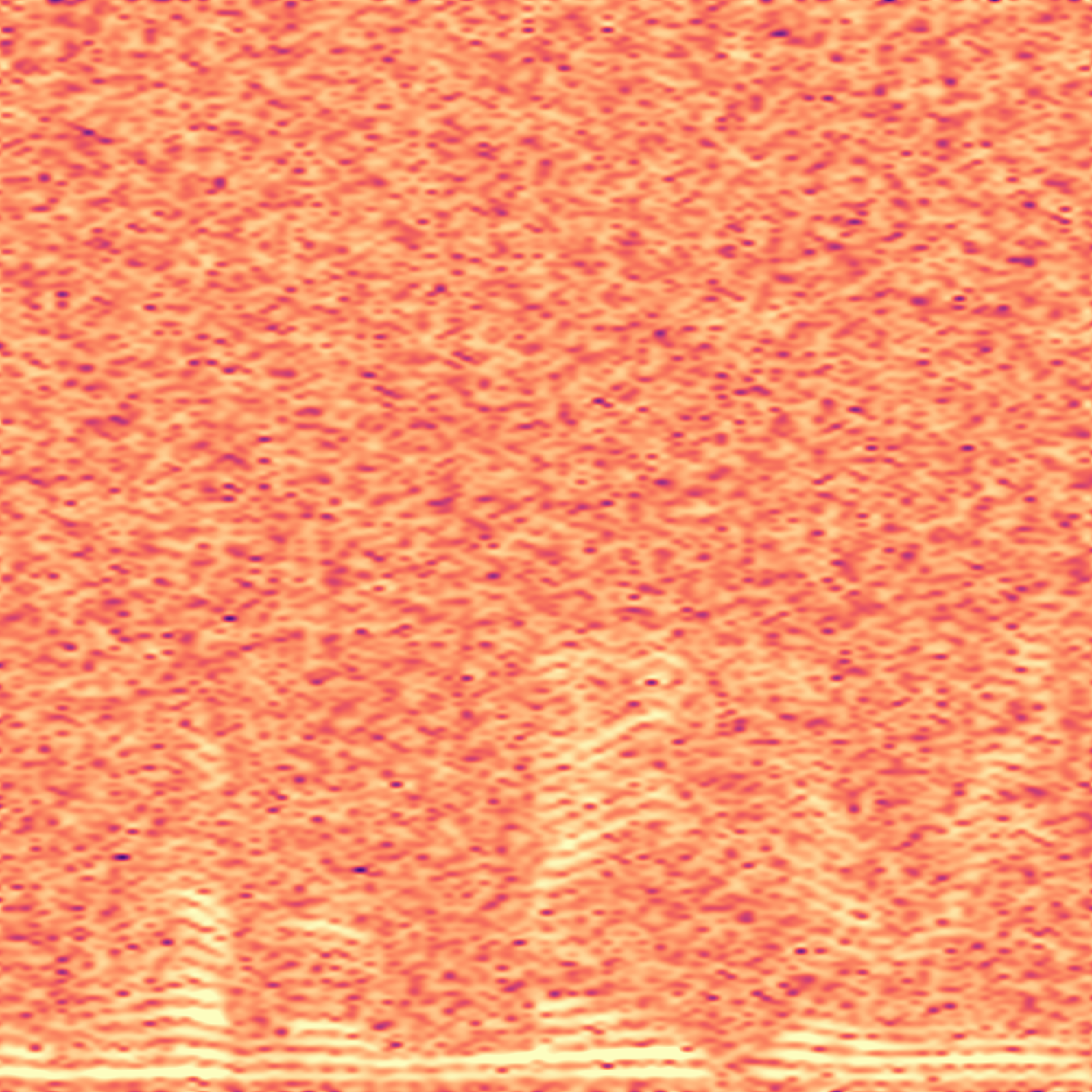}] (sNspec) {};
    \draw[-, dashed] (sN) to (sNspec);
    \node[mysquare, above=\sepnodeyspec of spsiN, anchor=south, fill overzoom image={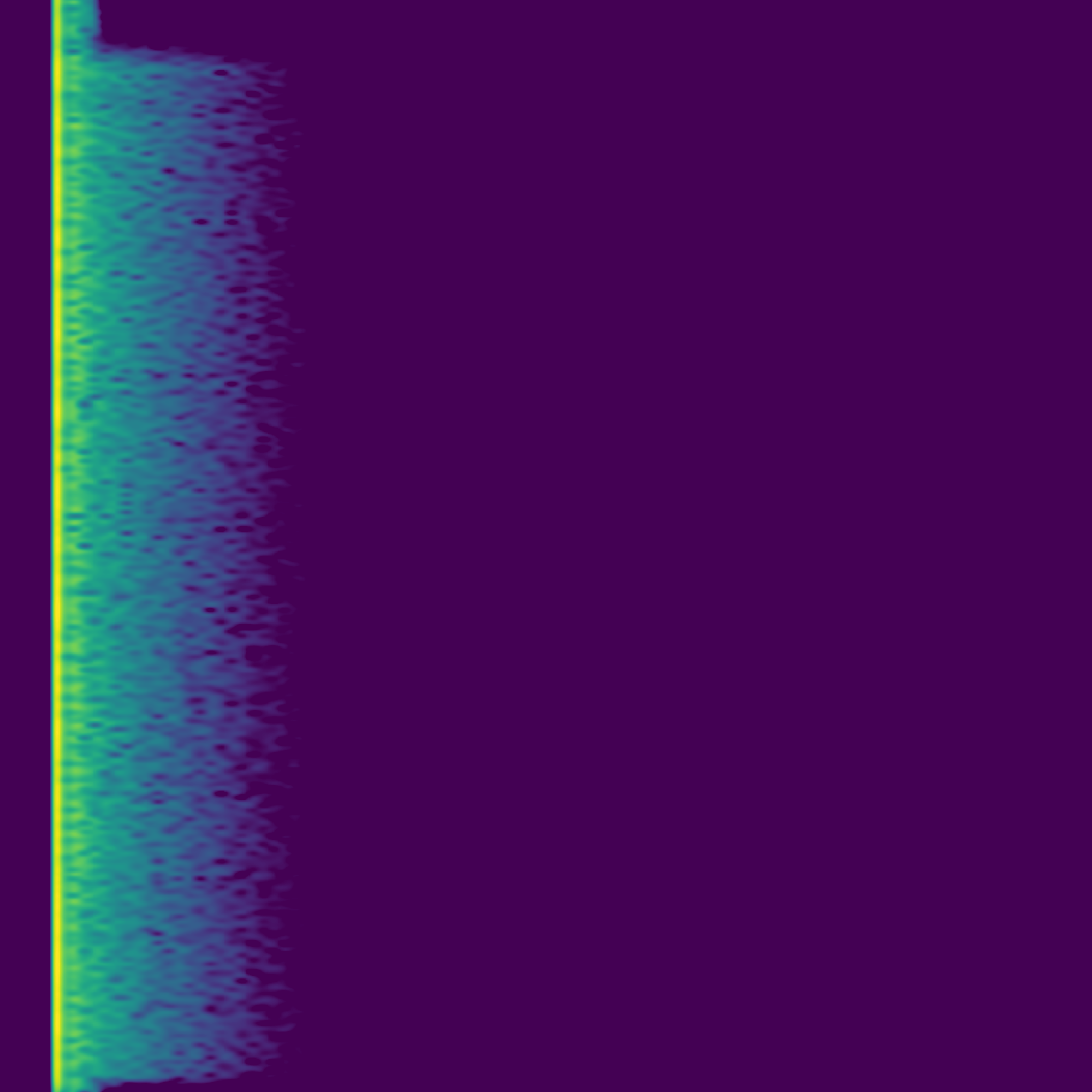}] (spsiNspec) {};
    \draw[-, dashed] (spsiN) to (spsiNspec);

    \node[mycircle, right=0.55cm of sN] (sn) {$\mathbf{x}_{n}$};
    \node[mycircle, right=0.55cm of spsiN] (spsin) {$\psi_{n}$};
    \draw[-, dotted] (sN.east) to (sn.west);
    \draw[-, dotted] (spsiN.east) to (spsin.west);
    \node[mysquare, below=\sepnodeyspec of sn, anchor=north, fill overzoom image={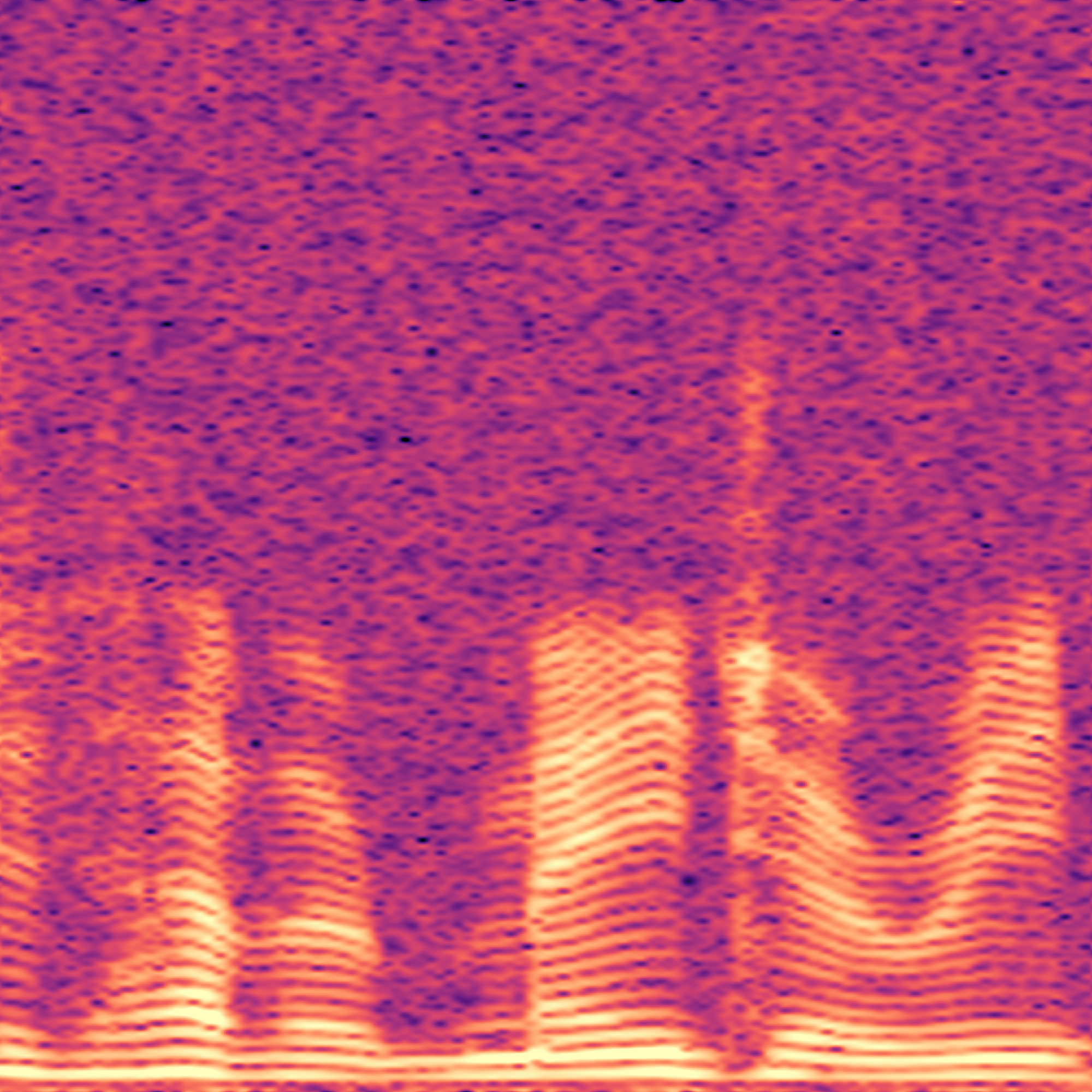}] (snspec) {};
    \draw[-, dashed] (sn) to (snspec);
    \node[mysquare, above=\sepnodeyspec of spsin, anchor=south, fill overzoom image={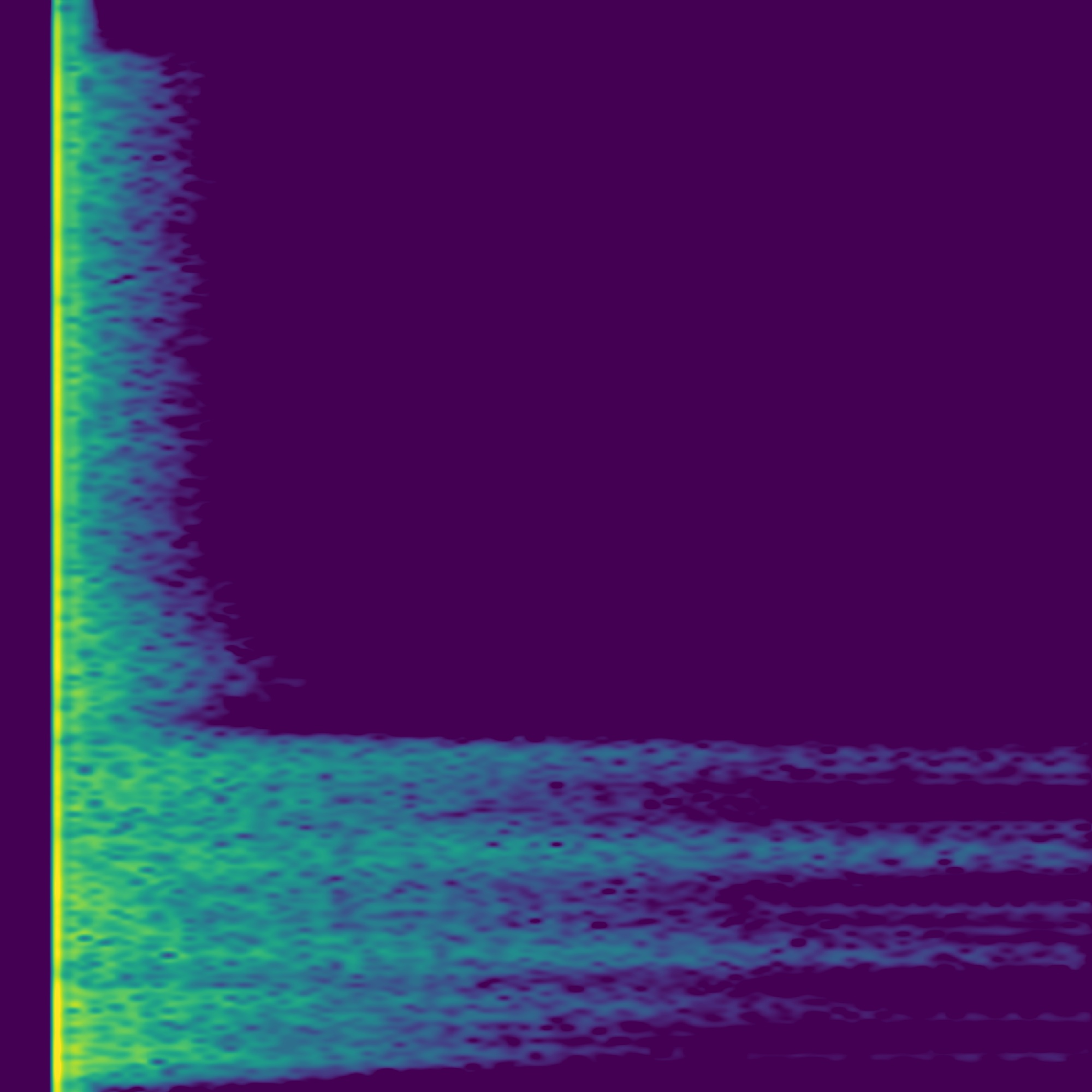}] (spsinspec) {};
    \draw[-, dashed] (spsin) to (spsinspec);
    
    \node[myrectangle, align=center, right=0.85cm of sn] (score)
    {Diffusion Prior
    \\ \vspace{-0.25cm}\\ \includegraphics[height=25pt]{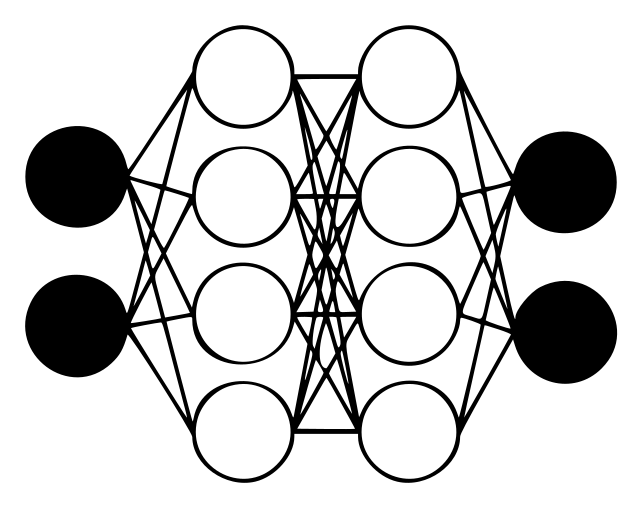}
    \vspace{-0.05cm}\\$p_{\theta^*}(\mathbf{x}_0)$
     };
    \node[right=0.5cm of score] (x0) {$\hat{\mathbf{x}}^n_0$};
    \draw[->] (sn) -- (score);
    \draw[->] (score) -- (x0);
    \node[myrectangle3, below=3.71cm of x0] (ll) {Likelihood \\$p_{\psi_n}(\mathbf{y} | \hat{\mathbf{x}}^n_0)$};
    \draw[->] (spsin) -- (ll);
    \draw[->] (x0) -- (ll);
    
    \node[mybranch, right=1.0cm of x0] (branchx0) {};
    \node[mybranch, right=0.455cm of ll] (branchpsi) {};
    \draw[-] (x0) -- (branchx0);
    \draw[-] (ll) -- (branchpsi);

    \node[myrectangle4, align=center, below=1.75cm of branchx0] (post) {Posterior \\$p_{\psi_n}( \hat{\mathbf{x}}^n_0 | \mathbf{y})$};
    \draw[->] (branchx0) -- (post);
    \draw[->] (branchpsi) -- (post);

    \node[myrectangle4, align=center, right=0.8cm of branchx0] (revdiff) {Reverse \\Diffusion};
    \node[myrectangle2, align=center, right=0.618cm of branchpsi] (riropti) {RIR\\Optimization};
    \node[at={([shift={(0,-0.25em)}]post.east)}] (out1) {};
    \node[at={([shift={(0,+0.25em)}]post.east)}] (out2) {};

    \draw[->] (branchx0) -- (revdiff);
    \draw[->] (branchpsi) -- (riropti);

    \draw[->] (out2) -| (revdiff.south);
    \draw[->] (out1) -| (riropti.north);

    \node[mycircle, right=8.5cm of sn] (snminus1)  {$\mathbf{x}_{n-1}$};
    \draw[->] (revdiff) -- (snminus1.west);
    \node[mycircle, right=8.5cm of spsin] (spsinminus1)  {$\psi_{n-1}$};
    \draw[->] (riropti) -- (spsinminus1.west);
    
    \node[mycircle, right=0.4cm of snminus1] (s0) {$\mathbf{x}_{0}$};
    \node[mycircle, right=0.4cm of spsinminus1] (spsi0) {$\psi_{0}$};
    \node[mysquare,  below=\sepnodeyspec of s0, anchor=north, fill overzoom image={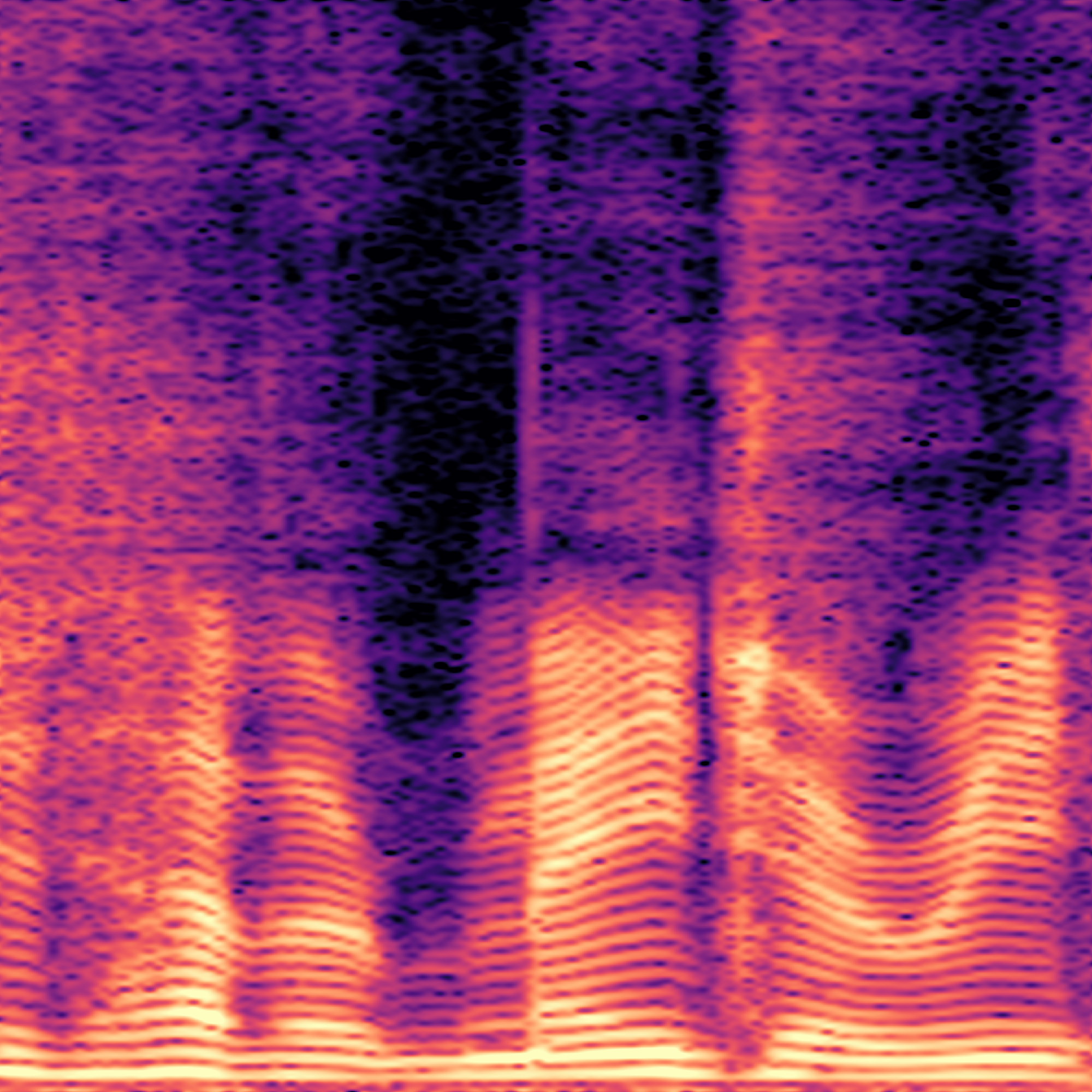}] (s0spec) {};
    \draw[-, dashed] (s0) to (s0spec);
    \node[mysquare,  above=\sepnodeyspec of spsi0, anchor=south, fill overzoom image={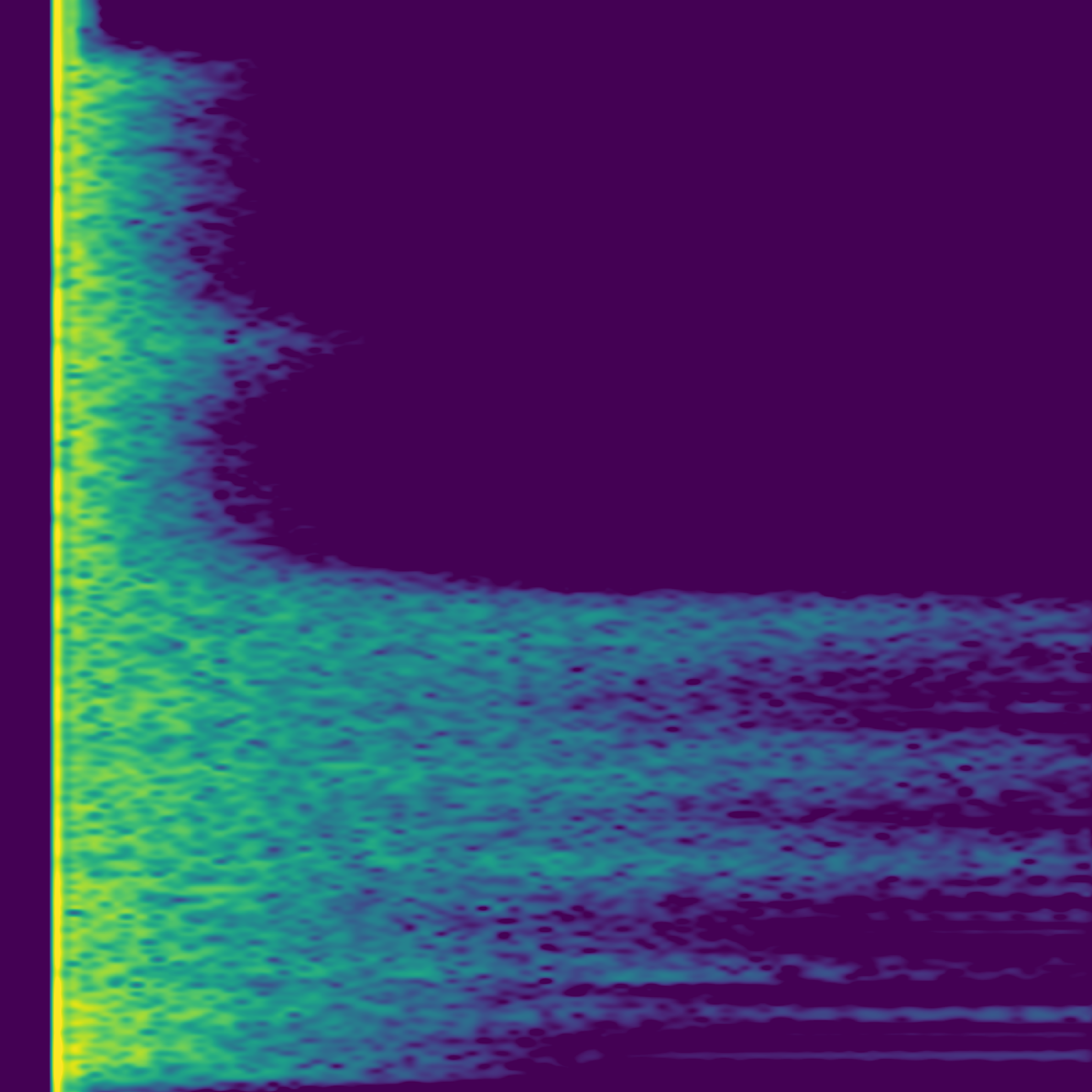}] (spsi0spec) {};
    \draw[-, dashed] (spsi0) to (spsi0spec);
    \draw[-, dotted] (snminus1.east) to (s0.west);
    \draw[-, dotted] (spsinminus1.east) to (spsi0.west);
    
\end{tikzpicture}~
\hspace{0.2cm}
\begin{tikzpicture}
\draw[-, dashed] (0, 1.2) to (0, -3.5);
\end{tikzpicture}~
\hspace{0.2cm}
\begin{tikzpicture}[scale=1.0, transform shape]

\node[rectangle, draw, minimum width=11.75em, minimum height=13.5em, fill=cb2!20!white] (rectangle_rir) at (0, 0.8) {};
\node[align=center, below=0.2cm of rectangle_rir.north] (rir_title) {RIR Optimization M-step};

\node[myrectangle_white, align=center, below=0.25 of rir_title] (rir_rec) 
{Rec. gradient step\\
$\nabla_{\psi_n} \, \mathcal{C}(\mathbf y, \mathcal{A}_{\psi_n}(\hat{\mathbf x}_0))$};
\node[myrectangle_white, align=center, below=0.25 of rir_rec] (rir_reg) {Noise regularization\\
$\nabla_{\psi_n} \mathcal{R}({\psi_n})$};
\node[myrectangle_white, align=center, below=0.25 of rir_reg] (rir_proj) {Constrain parameters};

\draw[->] (rir_rec) -- (rir_reg);
\draw[->] (rir_reg) -- (rir_proj);
\node[at={([shift={(-5.45em,-0.75em)}]rir_proj.south)}] (inter_south) {};
\node[above=0.3em of rir_rec.north] (inter_north) {};

\draw (rir_proj.south) |- node[pos=0.45, right] {$\times N_\mathrm{its.}$} (inter_south.center);
\draw (inter_south.center) |- (inter_north.center);
\draw[->] (inter_north.center) to (rir_rec.north);    
\end{tikzpicture}

\caption{\protect\textit{BUDDy: joint optimization alternating between RIR estimation and posterior sampling for speech reconstruction \cite{moliner2024buddy}.}}

\label{fig:diagram}

\end{figure*}

This section elaborates on the proposed method BUDDy, which extends the informed method presented in prior work \cite{lemercier2023derevdps} to the blind scenario, where the impulse response $\mathbf{h}$ is unknown. 
In Section~\ref{sec:operator}, we define a reverberation operator $\mathcal{A}_\psi(\cdot)$, which comprises a structured parametric model of the \ac{rir}, with parameters $\psi$. 
Section~\ref{sec:posterior-speech} then details the posterior sampling scheme used to obtain a speech utterance from the estimated posterior distribution.
Finally, Section~\ref{sec:rir-optimization} describes the optimization procedure for the reverberation model parameters $\psi$ using an \ac{em} formulation.
 The complete inference procedure is summarized in Algorithm \ref{alg:inference}, and an overview of the processing pipeline is given in Fig.\,\ref{fig:diagram}.

\subsection{Reverberation Operator}\label{sec:operator}

\subsubsection{Subband Filtering}

In contrast to \cite{lemercier2023derevdps} 
where the original time-domain convolution model is adopted, here we model reverberation using a subband filtering approximation in the \ac{stft} domain \cite{goodwin2009realization, avargel2007system}.
This approach enables us to incorporate prior knowledge about the characteristics of reverberation through a structured model of RIR magnitudes, characterized by exponential decays in each subband. By modeling reverberation in this way, we facilitate optimization and reduce the complexity of the RIR parameter search space.
Let $\mathbf{H} := \mathrm{STFT}(\mathbf{h}) \in \mathbb{C}^{N_\mathbf{h}\times K}$ represent the STFT of a RIR $\mathbf{h}$ with $N_\mathbf{h}$ time frames and $K$ frequency bins.
Similarly, let $\mathbf{X}\in \mathbb{C}^{M\times K}$, and $\mathbf{Y}
$, denote the STFTs of anechoic $\mathbf{x}_0$ and reverberant $\mathbf{y}$ speech signals, repectively.
The subband convolution operation applies independent convolutions along the time dimension of each frequency band:
\begin{equation}\label{eq:subband}
\mathbf{Y}_{m,k}=\sum_{n=0}^{N_h}\mathbf{H}_{n,k} \mathbf{X}_{m-n,k} \,.
\end{equation}
The resulting reverberant signal $\mathbf{Y} \in \mathbb{C}^{(M + N_\mathbf{h} - 1)\times K}$ can be transformed to time domain by applying the inverse STFT. 
The subband filtering model only approximates the time-domain convolution, as it does not account for the spectral leakage between frequency bands. 
However, it is empirically found to be a valid assumption in many scenarios involving reverberation \cite{avargel2007system, Nakatani2008a, Lemercier2023extending}.
Adding 50\% zero-padding to the end of the frames before computing the STFT is important to avoid cyclic convolution artifacts when transforming the resulting signal back to the time domain.

\subsubsection{RIR Prior}

In the blind scenario, estimating $\mathbf{H}$ is an ill-posed problem when the anechoic speech is unknown.
Therefore, we need to constrain the space of possible solutions by imposing a prior on $\mathbf{H}$.
We propose a structured, differentiable prior on $\mathbf{H}$, whose parameters $\psi$ can be estimated with gradient-based optimizers like Adam \cite{kingma2015adam}.
We denote the complete forward reverberation operator, including forward and inverse STFT operations, as 
$\mathcal{A}_\psi (\cdot): \mathbb{R}^{L} \rightarrow \mathbb{R}^{L+L_\mathbf{h}-1}$.
The whole processing pipeline is summarized in Algorithm \ref{alg:operator} with each component detailed below.

We denote as $\mathbf{A} \in \mathbb{R}^{N_\mathbf{h}\times K}$ and $\mathbf{\Phi} \in \mathbb{R}^{N_\mathbf{h}\times K}$ the RIR magnitudes and phases, respectively.
Following \cite{Habets2010}, we adopt an exponential decay model with learnable parameters controlling the decay time.
Since room materials exhibit frequency-dependent absorption behavior,
we parameterize the magnitude matrix $\mathbf{A}$ as a multi-band exponential decay model defined in $B<K$ frequency bands.
Let $\mathbf{A}^\prime \in \mathbb{R}^{N_\mathbf{h} \times B}$ be the subsampled version of $\mathbf{A}$ in the $B$ selected frequency bands.
Each frequency band $b$ is characterized by its weight $w_b$ and exponential decay rate $\alpha_b$, such that the corresponding subband magnitude filter is derived as
\begin{equation}
\mathbf{A}_{n,b}^\prime= w_{b} \cdot e^{-\alpha_b n } \,.
\end{equation}
Note that our parameterization can be extended to model coupled spaces by employing several decay parameters per band and summing their respective contributions \cite{Xiang2003evaluation}.
Once the parameters are estimated, we reconstruct the $K$-bands magnitudes $\mathbf{A}$ by interpolating the subsampled matrix $\mathbf{A}^\prime$ as $\mathbf{A}=\exp(\mathrm{lerp}(\log(\mathbf{A}^\prime)))$, where $\mathrm{lerp}$ represents linear interpolation on the frequency scale.
For this purpose, we employ the $\mathrm{torchcde}$ library, which facilitates efficient and differentiable interpolation \cite{kidger2020neuralcde}.
After interpolation of the magnitude matrix, we then obtain the time-frequency RIR $\mathbf{H}$ by multiplying the magnitude matrix $\mathbf{A}$ with the complex phase exponentials:
\begin{equation}
    \mathbf{H} = \mathbf{A} \odot e^{j \mathbf{\Phi}},
\end{equation}
where $j$ is the imaginary number and $\odot$ represents element-wise multiplication.
Given the general lack of phase structure, we optimize each phase factor in $\mathbf{\Phi}$ independently.
The RIR model $\psi = \{\mathbf{\Phi}, (w_b, \alpha_b)_{b=1,\dots, B} \}$ ultimately contains $2 \times B + N_\mathbf{h} \times K$ optimizable parameters.

\subsubsection{Projections}
We extend our forward reverberation operator with a series of projections to increase the likelihood of generating plausible RIRs. 
Thus, the time-frequency RIR $\mathbf{H}$ is further processed as
\begin{equation} \label{eq:projection}
    \overline{\mathbf{H}} = \mathrm{STFT} \left( \delta \oplus \mathcal{P}_\text{min}(\mathrm{iSTFT}(\mathbf{H})) \right) \,.
\end{equation}

This primarily ensures STFT consistency of $\overline{\mathbf{H}}$, exploiting the redundancy of the STFT representation and imposing inter-frame correlations between the RIR phases $\boldsymbol{\Phi}$.
We then enforce that the time-domain RIR estimate $\mathbf{h}$  has minimum-phase lag, using the Hilbert transform-based method in \cite{oppenheimschafer}.
This is indicated by the operator $\mathcal{P}_\text{min}$ and guarantees stability of the inverse RIR filter \cite{Miyoshi1988MINT}. We refer the reader to Appendix \ref{sec:appendix_minphase} for further details.
Finally, the operation $\delta \oplus ( \cdot )$ replaces the first sample of the time-domain \ac{rir} with a unit impulse.
This has the effect of injecting knowledge of the direct path in \(\overline{\mathbf{H}}\), and further requires us to correct the magnitude matrix $\mathbf{A}$ to account for this operation.
It is important to note that these steps are integral to the reverberation operator $\mathcal{A}_\psi(\cdot)$, which maps the parameters $\psi$ to the convolved signal $\mathcal{A}_\psi(\hat{\mathbf{x}}_0)$, as outlined in Algorithm \ref{alg:operator}. Since all operations are differentiable, we compute gradients with respect to $\psi$ by backpropagating through all operations. 
We propose a detailed ablation study of these projection and correction steps in Section \ref{sec:vctk:ablation_proj}.

\subsection{Posterior Speech Sampling} \label{sec:posterior-speech}

For sampling a speech utterance from the posterior distribution $p_\psi( \clean | \reverb)$, we adapt the posterior sampling algorithm of Section \ref{sec:dps} to the blind setting. 
As in Section \ref{sec:dps}, the pre-trained score model $\mathbf{s}_{\theta^\ast}(\mathbf{x}_\tau,\tau)$ is used to provide an implicit prior $p_{\theta^\ast} (\mathbf{x}_0)$ on anechoic speech with fixed parameters $\theta^\ast$.
The likelihood $p_\psi(\reverb | \clean)$ is approximated as
\begin{equation}\label{eq:likelihood_psi}
    p_\psi(\reverb | \clean) \propto \mathrm{exp} \left( - \zeta(\tau) 
\mathcal{C}(\reverb, \mathcal{A}_\psi(\tweedie) \right) \,,
\end{equation}
which is analogous to  \eqref{eq:likelihood}, but with the proposed reverberation operator $\mathcal{A}_\psi(\cdot)$ applied in place of a convolution with the oracle \ac{rir}, since the latter is unavailable in our blind scenario.
The sampling procedure then follows the ODE \eqref{eq:reverse-diff-estep} where the likelihood term $p_\psi(\reverb | \clean)$ defined in \eqref{eq:likelihood_psi} is used instead of the \ac{rir}-informed likelihood $p(\reverb | \clean)$ from \eqref{eq:likelihood}.

In order to guide and accelerate reverse diffusion, it is beneficial to use warm initialization, i.e., let the reverse diffusion process start from a speech sample $\mathbf{x}_{T}\sim \mathcal{N}(\mathbf{x}_\text{init}, \sigma^2(T) \mathbf{I})$, where $\mathbf{x}_\text{init}$ contains relevant information about the clean signal we wish to estimate.
Similar to \cite{saito2023unsupervised}, we obtain 
$\mathbf{x}_\text{init}$ through WPE \cite{Nakatani2008a}, a blind dereverberation algorithm based on variance-normalized delayed linear prediction. WPE performs mild dereverberation, which allows us to get closer to the clean speech, while not introducing much distortion to the signal.
As WPE is blind and unsupervised, our method remains fully blind and unsupervised as well.

\subsection{Reverberation Model Parameter Optimization} \label{sec:rir-optimization}

The posterior speech sampling approach described in Section \ref{sec:posterior-speech} relies on approximating the likelihood $p_\psi(\mathbf{y}|\mathbf{x}_0)$, where the main challenge is the dependence on the unknown parameters $\psi$. To estimate these parameters, we employ an \ac{em} formulation, drawing inspiration from \cite{nortier2023unsupervised, laroche2023fast}.
While the underlying algorithm follows a similar approach to prior works \cite{moliner2023zeroshot, moliner2024buddy}, the formalism has been refined to offer a more precise explanation of the method.
The optimization procedure maximizes the expected log-likelihood $\log p_\psi(\reverb | \clean)$ under the posterior distribution $p_\psi(\clean | \reverb)$
\begin{equation} \label{eq:em}
    \underset{\psi}{\mathrm{max}} \,
    \mathbb{E}_{p_\psi(\clean | \reverb)} \log p_\psi(\reverb | \clean) \,.
\end{equation}
This objective is unfortunately intractable, because the search quantity $\psi$ appears as a parameter of the distribution used to evaluate the expectation.
Therefore, we resort to an \ac{em} formulation alternating between an \emph{E-step} and an \emph{M-step}.
During the \emph{E-step}, the expectation is evaluated by drawing samples from the approximate posterior \( p_\psi(\mathbf{x}_0 | \mathbf{y}) \), given fixed RIR parameters \(\psi\). The \emph{M-step} then optimizes the RIR parameters \(\psi\) based on the clean speech estimates, maximizing the expected log-likelihood. Both steps are detailed below.

\subsubsection{E-Step} \label{sec:estep}

Given \ac{rir} parameters $\psi$, we wish to approximate the posterior speech distribution $p_\psi( \clean | \reverb) $ in order to evaluate the expectation in \eqref{eq:em}.
Drawing samples through our posterior sampling procedure explained above in Section \ref{sec:posterior-speech} would require running the entire diffusion process from $T$ to $T_\mathrm{min}$ at each \ac{em} iteration, as in \cite{nortier2023unsupervised}.
Instead, 
we follow \cite{laroche2023fast} and evaluate the expectation using the one-step denoised estimate
$\tweedie$ through \eqref{eq:tweedie} that is obtained at each reverse diffusion step.
Formally, the log-likelihood expectation is approximated as
\begin{equation}\label{eq:exp_approx}
    \mathbb{E}_{p_\psi(\clean | \reverb)} \log p_\psi(\reverb | \clean) \approx
     \mathbb{E}_{p(\clean | \state)} \log p_\psi(\reverb | \clean) \,,
\end{equation}
and the denoised posterior $p(\mathbf{x}_0 | \mathbf{x}_\tau)$ is modelled as a Dirac distribution located at the posterior mean $\tweedie$ \cite{chung_diffusion_2022}
\begin{equation}
    p(\clean | \state) \approx \delta(\tweedie) \,.
\end{equation}
This results in the following one-sample Monte Carlo estimate of the expectation \eqref{eq:em}
\begin{equation}
    \mathbb{E}_{p_\psi(\clean | \reverb)} \log p_\psi(\reverb | \clean) \approx
    \log p_\psi(\reverb | \tweedie) \,.
\end{equation}
This effectively integrates the M-step into the reverse diffusion, significantly accelerating inference.
Note that in \eqref{eq:exp_approx}, the dependency of $p_\psi(\clean | \reverb)$ to $\reverb$ is no longer explicit. 
However, this dependency still exists, as we notice that the Tweedie estimate $\tweedie$ is obtained from the current diffusion state $\state$, which has been itself sampled conditionally to $\reverb$ in \eqref{eq:reverse-diff-estep}.

\subsubsection{M-step} \label{sec:mstep}

Once the expectation in \eqref{eq:em} has been evaluated during the E-step, we can proceed to the M-step that will maximize the resulting objective. We add to the expected log-likelihood a regularization term facilitating the RIR parameter search during optimization
\begin{equation}\label{eq:noise_reg}
\mathcal{R}(\psi)=
\frac{1}{N_\mathbf{h}}
\rVert
 S_\mathrm{comp}(
\mathbf{h}_\psi)
-
S_\text{comp}(
\mathrm{sg} \left[ \mathbf{h}_\psi \right] +\nu(\tau) \mathbf{v} )\
\lVert_2^2 ,
\end{equation}
where $\mathbf{h}_\psi=\mathcal{A}_\psi(\delta)$ 
is the current time-domain RIR estimate, $\mathbf{v} \sim \mathcal{N}(\mathbf{0}, \mathbf{I})$ is a vector of white Gaussian noise, and $S_\mathrm{comp}$ is the magnitude compressed spectrogram, as defined in \eqref{eq:spec_comp}, with $N_\mathbf{h}$ time frames.
In the right-hand term, the ``stopgrad'' operator $\mathrm{sg}[ \cdot ]$ detaches the gradients of $\mathbf{h}_\psi$ from the optimization graph and Gaussian noise scaled by $\nu(\tau)$ is added.
We show in Appendix \ref{sec:appendix_noisereg} that this effectively injects multiplicative noise with standard deviation $\nu(\tau)$ in the \ac{rir} parameter gradients, taking inspiration from the ``regularization by denoising'' strategy used, e.g. in \cite{mardani2023variational} for guiding data reconstruction.
In our case, this regularization smoothes the RIR parameter optimization landscape by injecting small amounts of stochasticity during the optimization. 

The M-step finally consists in maximizing the resulting objective:
\begin{align}\label{eq:max}
    \psi & \leftarrow \underset{\psi}{\mathrm{arg max}}
    \left[ \log p_\psi(\reverb | \tweedie) - \mathcal{R}(\psi) \right] \nonumber \\
    &= \underset{\psi}{\mathrm{arg min}}
    \left[ \mathcal{C}(\reverb, \mathcal{A}_\psi(\tweedie)) + \mathcal{R}(\psi) \right] \,.
\end{align}
Since all operations, including our reverberation operator $\mathcal{A}_\psi(\cdot)$, are differentiable with respect to $\psi$, the M-step can be achieved with gradient-based optimizers, e.g. Adam \cite{kingma2015adam}.

During optimization, we further rescale the denoised speech estimate $\tweedie$ so that its root-mean-square power (RMS) matches the average RMS power of clean speech computed on the training set.
Using this additional constraint helps lift the indeterminacy when jointly optimizing the speech $\mathbf{x}_0$ and \ac{rir} parameters $\psi$.
This step is included in our ablation study in Section \ref{sec:vctk:ablation_proj}.
We also found it beneficial to constrain $w_b$ and $\alpha_b$ within a limited range to stabilize the optimization, specially at early stages. This is achieved by clamping the parameters to predefined minimum and maximum values after every optimization iteration, as specified in Appendix \ref{sec:appendix_exp_vctk:hp-op}.

\section{Experiments and Results} \label{sec:results}

In this section, we provide a comprehensive evaluation of BUDDy across various datasets and experimental setups. We detail the methodologies and baselines employed and present the results of our experiments.

\subsection{Speech Dereverberation} \label{sec:vctk}
We present dereverberation results on 16\,kHz speech data,
building upon the experiments conducted in prior work \cite{moliner2024buddy}.

\begin{table*}[t]
    \centering
    \caption{\centering\textit{Speech dereverberation results on reverberant VCTK datasets. 
    We indicate for each method in the table whether it is supervised or not. Boldface numbers indicate best performance for supervised and unsupervised methods separately. 
    }}
    \scalebox{0.79}{
    \begin{tabular}{l|c|cccc|cccc}
    
 \multicolumn{2}{c}{} & \multicolumn{4}{c}{VCTK-RealReverb (Matched)} &  \multicolumn{4}{c}{VCTK-SimulatedReverb (Mismatched)} \\
\cmidrule(lr){3-6} \cmidrule(lr){7-10}
Method& Unsup. & DNS-MOS & PESQ & ESTOI & SI-SDR & DNS-MOS & PESQ & ESTOI & SI-SDR \\

\midrule
\midrule
Reverberant & -
& 3.14 $\pm$ 0.52 & 1.61 $\pm$ 0.37 & 0.50 $\pm$ 0.14 & -12.3 $\pm$ 6.9
& 3.05 $\pm$ 0.47 & 1.57 $\pm$ 0.29 & 0.47 $\pm$ 0.11 & -12.5 $\pm$ 8.6 \\

\midrule \midrule

PSE 
& \xmark
& 3.75 $\pm$ 0.38 & 2.85 $\pm$ 0.55 & 0.80 $\pm$ 0.10 & 8.5 $\pm$ 6.5
& 3.61 $\pm$ 0.39 & 2.08 $\pm$ 0.47 & 0.64 $\pm$ 0.09 & \textbf{-8.4} $\pm$ \textbf{8.6} \\

SGMSE+M \cite{richter2023speech, Lemercier2022analysing} & \xmark
& 3.88 $\pm$ 0.32 & 2.99 $\pm$ 0.48 & 0.78 $\pm$ 0.09 & 0.2 $\pm$ 9.3
 & 3.74 $\pm$ 0.34 & 2.48 $\pm$ 0.47 & \textbf{0.69} $\pm$ \textbf{0.09} & \textbf{-8.4} $\pm$ \textbf{8.8} \\
 
StoRM \cite{Lemercier2022storm} & \xmark
& \textbf{3.90 $\pm$ 0.33} & \textbf{3.33 $\pm$ 0.48} & \textbf{0.82 $\pm$ 0.10} & \textbf{9.5 $\pm$ 6.5} 
& \textbf{3.83} $\pm$ \textbf{0.32} & \textbf{2.51} $\pm$ \textbf{0.53} & 0.67 $\pm$ 0.09 & -8.8 $\pm$ 10.2 \\ 
\midrule
\midrule

Yohena and Yatabe \cite{yohena2024single} & \cmark 
& 2.99 $\pm$ 0.56 & 1.80 $\pm$ 0.33 & 0.55 $\pm$ 0.12 &  -11.4 $\pm$ 7.7 
& 2.94 $\pm$ 0.44 & 1.71 $\pm$ 0.29 & 0.51 $\pm$ 0.10 &  -11.4 $\pm$ 8.6 \\

WPE \cite{Nakatani2008b} & \cmark 
& 3.24 $\pm$ 0.54 & 1.81 $\pm$ 0.42 & 0.57 $\pm$ 0.14 & -11.5 $\pm$ 8.2
& 3.10 $\pm$ 0.48 & 1.74 $\pm$ 0.37 & 0.54 $\pm$ 0.12 & -11.4 $\pm$ 8.8 \\
 
Saito et al. \cite{saito2023unsupervised} & \cmark 
& 3.22 $\pm$ 0.56 & 1.68 $\pm$ 0.40 & 0.51 $\pm$ 0.13 
& -11.7 $\pm$ 9.2 
& 3.12 $\pm$ 0.52 & 1.70 $\pm$ 0.33 & 0.52 $\pm$ 0.10 
& -11.7 $\pm$ 8.5 
\\

GibbsDDRM \cite{murata_gibbsddrm_2023} & \cmark
& 3.33 $\pm$ 0.53 & 1.70 $\pm$ 0.37 & 0.51 $\pm$ 0.13
& -11.9 $\pm$ 8.5 
& 3.30 $\pm$ 0.52 & 1.75 $\pm$ 0.36 & 0.52 $\pm$ 0.11
& -11.8 $\pm$ 8.9 
\\

RVAE-EM \cite{Wanf2024rvaeem} & \cmark
& 3.05 $\pm$ 0.53 & 1.83 $\pm$ 0.32 & 0.54 $\pm$ 0.11 & -12.2 $\pm$ 8.1
& 3.00 $\pm$ 0.45 & 1.76 $\pm$ 0.30 & 0.52 $\pm$ 0.10 & -11.8 $\pm$ 8.5 \\

BUDDy (ours)
& \cmark 
& \textbf{3.76} $\pm$ \textbf{0.41} & \textbf{2.30} $\pm$ \textbf{0.53} & \textbf{0.66} $\pm$ \textbf{0.12} & 
\textbf{-7.8} $\pm$ \textbf{8.40} 
& \textbf{3.74} $\pm$ \textbf{0.38} & \textbf{2.24} $\pm$ \textbf{0.54} & \textbf{0.65} $\pm$ \textbf{0.12} & \textbf{-8.4} $\pm$ \textbf{9.9}  \\

\midrule
    \bottomrule
    \end{tabular}
    }
    \label{tab:blind_results}
\end{table*}

\subsubsection{Data}
\label{sec:vctk:data}

We use VCTK 
\cite{vctkdataset} 
as clean speech, selecting 103 speakers for training, two for validation, and two for testing. 
The total dataset represents 44\,h of audio, which we down-sample to 16\,kHz for our experiments.
We curate \acp{rir} from various public datasets \cite{prawda2022calibrating, Eaton2015TheAC, Jeub2009ABR, fejgin2023brudex, datasetPalimpsest, kearney2022bbc, datasetBUT, dietzen2023myriad, traer2016mit}. 
In total we approximately obtain 10k RIRs, and split them between training, validation and testing using ratios 0.9/0.05/0.05.
We use the three following test benchmarks for our experiments:

\begin{itemize}
    \item \textit{VCTK-RealReverb}: This test set matches the training speech corpus and reverberant conditions. It contains 500 speech sequences from the two VCTK speakers $\mathrm{p226}$ and $\mathrm{p287}$ reserved for testing. These utterances are convolved with the curated RIRs reserved for testing, which are therefore excluded from the paired reverberant/anechoic dataset used to train supervised approaches (see next Section \ref{sec:vctk:baselines}).

    \item \textit{VCTK-SimulatedReverb}: This test set uses the same test speech utterances as VCTK-RealReverb, matching the training speech corpus. However, the reverberant conditions are obtained by simulating \acp{rir} with \texttt{pyroomacoustics}\cite{Scheibler2018PyRoom}. 
    For ease of comparison, we choose simulation parameters such that the distributions of reverberation times and direct-to-reverberation ratios of the simulated mismatched dataset approximately match those of the matched dataset using real \acp{rir}.

    \item \textit{DDS-DAPS-RealRecorded}: This test benchmark is a subset of the DDS dataset \cite{Li2021DDS}, which records utterances in real rooms, as opposed to the convolution model used so far. The resulting speech contains natural reverberation and background noise (e.g., air conditioning, device noise). 
    For this benchmark, we selected two of the most reverberant rooms from the dataset, $\mathrm{confroom1}$ and $\mathrm{confroom2}$ along with four microphone positions located at distances greater than 1 m from the source. 
    We use the portion containing utterances from 20 speakers from the DAPS corpus \cite{Mysore2014DAPS}, therefore providing a test benchmark which completely mismatches the training conditions in terms of speech corpora and reverberation conditions. We downsample all utterances to 16\,kHz for fair comparison.
    
\end{itemize}

\subsubsection{Baselines}
\label{sec:vctk:baselines}

We compare our method BUDDy to several blind supervised baselines such as the predictive approach in  \cite{Lemercier2022analysing}, which will denote as PSE in the following (for \textit{predictive speech enhancement}), and diffusion-based SGMSE+ \cite{richter2023speech} and StoRM \cite{Lemercier2022storm}.
The STFT-based diffusion model in SGMSE+ and StoRM uses supervision in both the network conditioning and the diffusion trajectory parameterization; PSE uses a classical $L^2$-distance between the clean target and its estimate and has virtually the same architecture as SGMSE+. 
These methods require coupled reverberant/anechoic speech, which we generate using our curated RIR and anechoic speech datasets.
The reverberant speech is obtained by first aligning the direct path of the RIR to its first sample, then convolving the anechoic speech from VCTK with the resulting RIR, and finally normalizing it to reach the same loudness \cite{ituloudness}
as the anechoic speech.

We also include blind unsupervised approaches leveraging traditional methods such as WPE \cite{Nakatani2008a} and Yohena and Yatabe \cite{yohena2024single}, as well as generative models
Saito et al. \cite{saito2023unsupervised}, GibbsDDRM \cite{murata_gibbsddrm_2023} and RVAE-EM \cite{Wanf2024rvaeem}.
Please see Appendix \ref{sec:appendix_exp_vctk:baselines} for more details on baselines.

\subsubsection{Hyperparameters}
\label{sec:vctk:hyperparameters}

As in \cite{moliner2024buddy, lemercier2023derevdps}, we implement the unconditional score model architecture with NCSN++M\cite{Lemercier2022analysing,Lemercier2022storm}, which is a convolution-based neural network operating in the complex STFT domain.
NCSN++M is also used as the base architecture for PSE, SGMSE+ and StoRM.
Details on the architecture, training configuration, reverberation operator and diffusion hyperparameters can be found in appendices \ref{sec:appendix_exp_vctk:hp}., \ref{sec:appendix_exp_vctk:hp-op} and \ref{sec:appendix_exp_vctk:hp-diff}, respectively.

\subsubsection{Instrumental metrics}
\label{sec:vctk:metrics}

For instrumental evaluation of the speech dereverberation performance, we use the intrusive \ac{pesq} \cite{Rix2001PESQ} and \ac{estoi} \cite{Jensen2016ESTOI} for assessment of speech quality and intelligibility respectively. We also use the non-intrusive DNS-MOS \cite{reddy2021dnsmos}, a DNN-based \ac{mos} approximation following the ITU-T P.835 recommendation \cite{itu835}.
We also report SI-SDR \cite{Leroux2019SISDR}, however it must be noted that such point-wise distance metrics do not represent well the performance of generative models, given the natural variability of the corresponding estimates.

\subsubsection{Instrumental evaluation results}
\label{sec:vctk:results}

We display in Table \ref{tab:blind_results} the dereverberation results for all blind methods, both supervised and unsupervised.
Blind supervised approaches PSE, SGMSE+ and StoRM generally perform better than unsupervised methods as they benefit from supervision at training time.
However, we can observe the limited generalization ability of supervised approaches on the VCTK-SimulatedReverb when reverberant conditions are not the same as those presented during training.
Our method BUDDy, however, seamlessly adapts to changing acoustics since it was trained without supervision. This enables BUDDy to retain its performance from VCTK-RealReverb to VCTK-SimulatedReverb, where supervised methods like PSE lose up to 0.77 PESQ points in mismatched reverberant conditions.

When evaluating on the DDS-DAPS-RealRecorded dataset, as reported in Table \ref{tab:dds_results}, the generalization gap between our method and the supervised baselines increases even further. On this benchmark, BUDDy outperforms the best supervised baselines SGMSE+ and StoRM on DNS-MOS and ESTOI and has similar PESQ scores. This experiment highlights several points. First, BUDDy is naturally robust to background noise although it is not present in its signal model, echoing the conclusions of our prior work \cite{lemercier2023derevdps}. Informal listening suggests that the background noise is left untouched by the algorithm, and therefore it can be easily dealt with after processing.
Furthermore, BUDDy's unconditional diffusion model trained on the anechoic VCTK dataset seems to generalize well to DAPS speech utterances.
Finally, BUDDy also performs strong dereverberation when handling a realistic reverberant model, since the utterances in the DDS-DAPS-RealRecorded dataset are directly recorded in a reverberant room, and not produced by a convolution between an utterance and a measured \ac{rir}. 
On the contrary, the supervised baselines suffer from the multiple mismatches between the presented benchmark and the training conditions, be it with regard to background noise, different speech corpora or reverberant model. 
This strengthens the position of BUDDy as a robust unsupervised baseline versus top-performing supervised baselines.

Overall, BUDDy performs far better than all other blind unsupervised baselines. For instance, BUDDy outperforms RVAE-EM by as much as 0.47 PESQ and 0.12 ESTOI points.
Indeed, traditional unsupervised methods \cite{Nakatani2008a, yohena2024single}
only draw limited benefits from their uninformed Gaussian prior on anechoic speech, while 
diffusion-based Saito et al. \cite{saito2023unsupervised} and GibbsDDRM \cite{murata_gibbsddrm_2023} seem to only marginally deviate from their WPE initialization.
RVAE-EM \cite{Wanf2024rvaeem} also obtains low instrumental scores, but informal listening suggested that its dereverberation abilities were superior to those of WPE.

\begin{table}[t]
    \centering
    \caption{\centering\textit{Speech dereverberation results on DDS-DAPS-RealRecorded  reverberant benchmark. 
    Boldface numbers indicate best performance for supervised and unsupervised methods separately.
    Underlined numbers indicate best performance across all methods.
    }}
    \scalebox{0.79}{
    \begin{tabular}{l|c|ccc}
    
Method& Unsup. & DNS-MOS & PESQ & ESTOI
\\

\midrule
\midrule
Reverberant & -
 & 2.35 $\pm$ 0.59 & 1.30 $\pm$ 0.16 & 0.63 $\pm$ 0.10
 \\
 \midrule \midrule

PSE 
& \xmark
& 2.91 $\pm$ 0.55 & 1.65 $\pm$ 0.40 & 0.74 $\pm$ 0.14
\\

SGMSE+M \cite{richter2023speech} & \xmark
 & 3.21 $\pm$ 0.52 & \underline{\textbf{2.14} $\pm$ \textbf{0.42}} & \textbf{0.83} $\pm$ \textbf{0.09}
 \\
 
StoRM \cite{Lemercier2022storm} & \xmark
& \textbf{3.48} $\pm$ \textbf{0.45} & 2.12 $\pm$ 0.50 & 0.82 $\pm$ 0.12
\\
\midrule
\midrule

WPE \cite{Nakatani2008b} & \cmark 
& 2.64 $\pm$ 0.59 & 1.48 $\pm$ 0.29 & 0.70 $\pm$ 0.11
\\

RVAE-EM \cite{Wanf2024rvaeem} & \cmark
& 2.68 $\pm$ 0.55 & 1.59 $\pm$ 0.33 & 0.71 $\pm$ 0.11
\\

BUDDy (ours)
& \cmark 
& \underline{\textbf{3.55}$\pm$ \textbf{0.49}} & \textbf{2.11} $\pm$ \textbf{0.47} & \underline{\textbf{0.86} $\pm$ \textbf{0.10}} 
\\

\midrule
    \bottomrule
    \end{tabular}
    }
    \label{tab:dds_results}
\end{table}

\subsubsection{Ablation study}
\label{sec:vctk:ablation_proj}

We conduct an ablation study to evaluate the impact of the projection step \eqref{eq:projection} introduced in the operator  optimization (see Section~\ref{sec:operator}).
We present the results in Table~\ref{tab:ablation_proj} and observe that, although the minimum-phase consistency projection has a theoretical justification as a mean to enhance the stability of the inverse RIR during optimization, its practical effect appears negligible, which can be due to a mismatch with the fact that real RIRs are generally mixed-phase filters \cite{Neely1979Invertibility}.
However, we observe that the other operations in the projection step, i.e. STFT consistency, enforcement of the direct path, and speech magnitude constraint, are all instrumental in guiding BUDDy toward a solution with higher fidelity to clean speech, as measured by PESQ.
We show DNS-MOS figures out of completeness. However,  DNS-MOS variations are small across ablations and not indicative of fidelity to reference speech as DNS-MOS is not intrusive.

Additionally, we examine the effect of parameterizing the likelihood model with a $L^2$-distance on compressed spectrograms rather than on waveforms as in previous work \cite{lemercier2023derevdps}.
To do so, we replace the cost function $C(\cdot, \cdot)$ from \eqref{eq:objective}, which is based on compressed \ac{stft} representations, with a simpler waveform-domain $L^2$-distance, and we empirically pick the optimal corresponding scaling factor $\tilde{\zeta}$.
The results clearly show the superiority of the proposed cost function \eqref{eq:objective} using compressed \ac{stft} representations.

\subsubsection{Listening experiment}
\label{sec:vctk:listening_test}

Instrumental metrics offer only limited insights into the performance of dereverberation algorithms \cite{Goetze2014}.
We therefore conduct a listening experiment based on the MUSHRA recommendation \cite{ITURmushra} to assess the performance of BUDDy as perceived by human listeners.
The test comprised 12 pages, featuring 6 reverberant speech utterances from the  \textit{VCTK-RealReverb} (matched) and the \textit{VCTK-SimulatedReverb} (mismatched) sets.
Participants were asked to rate the different stimuli with a single number representing overall quality, taking into account factors such as voice distortion, residual reverberation, and potential artifacts 
\cite{Goetze2014}. 
The test stimuli include our proposed method BUDDy, the unsupervised WPE \cite{Nakatani2008a} and RVAE-EM \cite{Wanf2024rvaeem}, as well as the supervised baselines PSE 
and SGMSE \cite{richter2023speech},
Further details on the organization of the listening experiment are reported in Appendix \ref{sec:appendix_exp_vctk:listening}.

\begin{table}[t]
    \centering
    \caption{\centering\textit{Ablation study on 
    VCTK-RealReverb.
    }}
    \resizebox{\columnwidth}{!}{%
    \begin{tabular}{l|cc}
    Method 
    & PESQ
    & DNS-MOS 
    \\
    
\midrule
\midrule
Reverberant 
& 1.61 $\pm$ 0.37
& 3.14 $\pm$ 0.52 
\\
\midrule

BUDDy
& \textbf{2.30 $\pm$ 0.53}
& 3.76 $\pm$ 0.41 
\\

- Minimum-phase Consistency
& \textbf{2.30 $\pm$ 0.57}
& 3.81 $\pm$ 0.40 
\\

$\;$ - RMS Power Constraint
& 2.22 $\pm$ 0.50
& 3.64 $\pm$ 0.50 
\\

$\;$ $\;$ - Fixed Direct Path
 & 2.10 $\pm$ 0.46
 & 3.78 $\pm$ 0.44 
 \\

$\;$ $\;$ $\;$ - STFT Consistency
& 1.96 $\pm$ 0.41
& \textbf{3.84 $\pm$ 0.39} 
\\
\midrule

$L^2$-Distance for $\mathcal{C}(\cdot,\cdot)$  
& 1.86 $\pm$ 0.47
& 3.36 $\pm$ 0.56 
\\

\midrule
    \bottomrule
    \end{tabular}
    }
    \label{tab:ablation_proj}
\end{table}

\begin{figure}[t]
    \centering
    \includegraphics[width=\columnwidth]{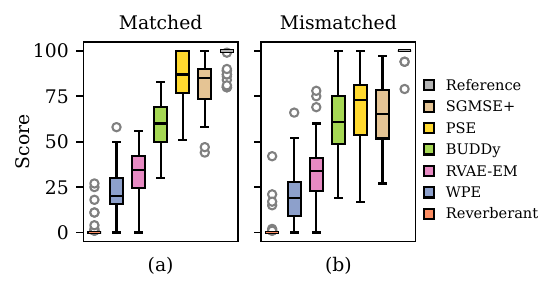}
    \caption{\protect\centering\textit{Listening test results on VCTK-RealReverb and VCTK-SimulatedReverb.
    The boxplot shows first quartile, median, and third quartile.
    }}
    \label{fig:listening_test_results}
\end{figure}

The results of the experiment are presented in Fig.~\ref{fig:listening_test_results}. 
It can be observed that the unsupervised baselines WPE and RVAE-EM received low scores. Yet, RVAE-EM performs consistently better than WPE in this listening experiment, as opposed to what is suggested by instrumental metrics in Table~\ref{tab:blind_results}.
In the matched test set (Fig.~\ref{fig:listening_test_results}a), BUDDy obtained significantly lower scores than PSE and SGMSE+ ($p < 0.001$ in a paired Welch test). However, in the mismatched set, PSE and SGMSE+ suffered a decrease in performance, losing up to $20$ points (out of $100$), while BUDDy mantained similar scores.
In that case, there is no significant difference in performance between the three approaches ($p> 0.1$), which closes the gap between
BUDDy and the top-performing supervised baselines in this mismatched setting, highlighting the advantage provided by unsupervised learning.

\subsubsection{Computational complexity}

A current limitation of the proposed method lies in its high computational budget. Compared to classical diffusion approaches, such as SGMSE+ \cite{richter2023speech}, our method requires several optimization iterations for the reverberation operator parameters at each reverse diffusion step, which increases the per-step inference time from 0.03s to 0.10s between SGMSE+ and BUDDy for each second of speech signal processed, as measured on a NVIDIA RTX A6000 GPU.
Together with the larger number of reverse diffusion steps, the overall inference time is significantly higher than e.g. SGMSE+ or predictive approaches like PSE in Table~\ref{tab:blind_results}.
Thus, while BUDDy is able to produce unprecedented high quality speech dereverberation without training on paired data, it cannot currently do so without an increase in computational complexity.

\subsection{Singing Voice Dereverberation} \label{sec:nhss}

We extend our evaluation benchmark to include the related task of singing voice dereverberation.

\begin{table}[t]
        \caption{\centering\textit{Singing voice dereverberation results on NHSS dataset. 
    Boldface numbers indicate best performance for supervised and
unsupervised methods separately
    }}
    \resizebox{\columnwidth}{!}{
    \begin{tabular}{l|c|cc|cc}
    
 \multicolumn{2}{c}{} & \multicolumn{2}{c}{Matched} &  \multicolumn{2}{c}{Mismatched} \\
\cmidrule(lr){3-4} \cmidrule(lr){5-6}

Method & Unsup. &  $\ell^1$ STFT & FAD &  $\ell^1$ STFT & FAD  \\

\midrule
\midrule
Reverberant & -
&  1.98 $\pm$ 0.66 & 6.41
& 1.86 $\pm$ 0.56 & 5.65
\\
\midrule \midrule

PSE  & \xmark & 1.56 $\pm$ 0.70 & 1.15
& 2.24 $\pm$ 0.78 & 1.79
\\
SGMSE+ \cite{richter2023speech}  & \xmark &  \textbf{1.32 $\pm$ 0.51} & \textbf{0.82}
& \textbf{1.37 $\pm$ 0.39} & \textbf{0.65}
\\ 
\midrule
\midrule

WPE \cite{Nakatani2008a} & \cmark
&  2.02 $\pm$ 0.65 &  4.66
& 2.29 $\pm$ 0.66 & 5.74
\\
 
 Saito et al. \cite{saito2023unsupervised} & \cmark
 & 1.95 $\pm$ 0.65 & 5.46
 & \textbf{1.77 $\pm$ 0.52} & 4.90
 \\
BUDDy 
& \cmark
 & \textbf{1.90 $\pm$ 0.59} & \textbf{0.88}
 & 1.91 $\pm$ 0.50 & \textbf{0.60}
\\

\midrule
    \bottomrule
    \end{tabular}
    }
    \label{tab:nhss}
\end{table}

\begin{figure}[t]
    \centering
    \includegraphics[width=\columnwidth]{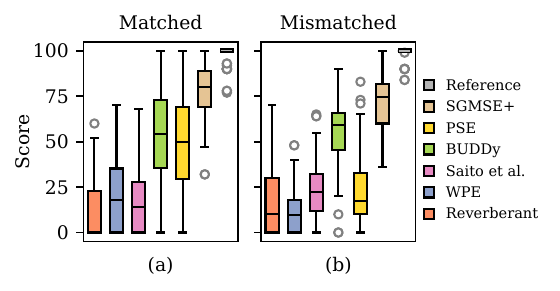}
    \caption{\protect\centering\textit{Listening test results on singing voice datasets NHSS-RealReverb and NHSS-SimulatedReverb. 
    The boxplot shows first quartile, median, and third quartile.
    }}
    \label{fig:sv_listening_test_results}
\end{figure}

\subsubsection{Data}
\label{sec:nhss:data}

We collect several publicly available singing voice datasets \cite{huang2021multi, wang2022opencpop,  zhang2022m4singer, duan2013nus, choi2020children, koguchi2020pjs}.
These datasets feature over 94 h of studio-quality solo singing from a diverse array of singers and singing styles, spanning various languages. The majority of the recordings are in Chinese, followed by English, Japanese, and Korean.
All datasets are down-sampled to 44.1\,kHz.
For testing, similar to \cite{murata_gibbsddrm_2023}, we use the sung part of NHSS \cite{sharma2020nhss, sharma2019combination}. The NHSS dataset contains 100 English-language pop songs, 10 for each of the five male and five female singers recruited.
We select a subset (90\%) of the RIRs curated for the VCTK-based experiments, such that we only retain the RIRs whose original sample rate is at least 44.1\,kHz. The resulting reverberant test set is referred to as \textit{NHSS-RealReverb}.
As for the speech voice experiments, we also prepare mismatched test set denoted as \textit{NHSS-SimulatedReverb} using simulated RIRs sampled at 44.1\,kHz.

\subsubsection{Baselines}

We evaluate the performance of BUDDy against two unsupervised baselines: WPE \cite{Nakatani2008a} and the unsupervised method from Saito et al. \cite{saito2023unsupervised} which was originally designed for singing voice dereverberation. 
Additionally, we train supervised baselines following the same approach as in the speech dereverberation experiments described in Section \ref{sec:vctk:baselines}. Specifically, we compare BUDDy to PSE and SGMSE+, which share the same architectural design as BUDDy’s diffusion model, as detailed in Section \ref{sec:nhss:hyperparameters}.

\subsubsection{Hyperparameters}
\label{sec:nhss:hyperparameters}

Similar to the speech dereverberation experiments, we adopt NCSN++ as the score model architecture. However, we adjust its hyperparameters to accommodate the higher sampling rate of 44.1 kHz. 
More details concerning the architecture and specific training configuration and inference hyperparameters are reported in Appendix \ref{sec:appendix_exp_nhss}

\subsubsection{Evaluation metrics}
\label{sec:nhss:metrics}

Objective metrics for evaluating singing voice restoration tasks are limited compared to those available for speech processing.
Following \cite{saito2023unsupervised}, we use the $\ell^1$-distance in the magnitude STFT domain and a Fréchet Audio Distance (FAD) using a VGGish embedding \cite{kilgour2018fr}. 
However, these are only limited in interpretability and hardly relate to listening impression \cite{kandpal2022music,gui2024adapting}.
Therefore, we complete this evaluation benchmark with a listening test with 10 participants, using a similar setup as reported in Section \ref{sec:vctk:listening_test}. 
The test included 12 reverberant singing voice examples from the NHSS dataset, containing 6 reverberant singing voice examples from each of the matched and mismatched datasets.
The instructions were identical to those reported in Appendix \ref{sec:appendix_exp_vctk:listening}.

\subsubsection{Results}
\label{sec:nhss:results}

The results from the instrumental evaluation are reported in Table~\ref{tab:nhss} and those from the listening test in Fig.~\ref{fig:sv_listening_test_results}.
The results show that BUDDy largely outperforms the unsupervised baselines and the supervised baseline PSE on FAD and subjective listening. In particular, in the mismatched test set, PSE fails to produce high-quality speech whereas BUDDy remains consistent with its good performance on the matched test set.
BUDDy outperforms SGMSE+ on FAD in the mismatched setting, However, although it loses a few points in the mismatched scenario, SGMSE+ surpasses BUDDy by a small yet statistically significant margin on both test benchmarks in the listening experiment.

\subsection{Robustness of RIR-Informed Methods} \label{sec:robustness}

Informed dereverberation algorithms such as \cite{Kodrasi2014, lemercier2023derevdps} assume complete knowledge of the room acoustics as provided by the RIR $\mathbf{h}$. However, as pointed out in Section~\ref{sec:introduction}, even if the RIR is perfectly known, single channel dereverberation is not trivial as RIRs are mixed-phase systems, such that causal and stable inverse filters do not exist \cite{Neely1979Invertibility}.
We examine here the sensitivity of informed dereverberation methods in \textit{partially blind} scenarios, i.e. when RIRs are known up to estimation errors.

\vspace{0.5em}
\subsubsection{Baselines}

We include several informed dereverberation approaches for comparison.
The baseline InfDerevDPS is inspired from previous work \cite{lemercier2023derevdps}
and fully described in Section \ref{sec:dps}.
We also include the regularized inverse filtering method RIF+Post \cite{Kodrasi2014}.
The second method RIF+Post \cite{Kodrasi2014} performs regularized inverse filtering in the Fourier domain, followed by traditional speech enhancement \cite{Breithaupt2008novel}. More details on these baselines can be found in Appendix~\ref{sec:appendix_exp_vctk:informed_baselines}.

\vspace{0.5em}
\subsubsection{Synthetic RIR estimation errors}

We first study the case where the oracle RIR is corrupted by Gaussian noise.
The results displayed in Fig.~\ref{fig:robustness} indicate that the performance of both the diffusion-based and the traditional method dwindles as the noise power increases.
This suggests that informed methods have very limited robustness to errors in the provided RIR. This is a crippling drawback since obtaining a perfect sample-wise estimation of RIRs is an arduous problem, given their stochastic nature \cite{Habets2010}.

\vspace{0.5em}
\subsubsection{DNN-based RIR estimation errors}

We now shift to a more realistic scenario where the RIR is blindly estimated from the reverberant speech by a DNN, 
since in practice RIR estimation errors are unlikely to be perfectly Gaussian-distributed.
In particular, we employ FiNS \cite{steinmetz2021filtered}, a state-of-the-art supervised RIR estimator which obtains RIR estimates based on the reverberant utterance (see Section \ref{sec:rir:baseline} for details).
We compare in Table \ref{tab:inf_vs_blind_results} results where the RIR is perfectly known (i.e. informed scenario) versus when it is estimated by FiNS (i.e. partially blind).
The acoustic conditions in the considered evaluation set match those of the training set. Therefore, since FiNS was trained in a supervised fashion using paired reverberant/RIR data, it is expected to perform well on such conditions.
The dereverberation performance of both InfDerevDPS and RIF+Post is very poor when the \ac{rir} is estimated with FiNS \cite{steinmetz2021filtered}, as opposed to when the RIR is perfectly known.
Yet, through informal listening, we notice that FiNS produces perceptually reasonable RIR estimates, which highlights the very limited robustness of informed methods when estimation errors, even imperceptible, affect the RIR knowledge.
This all suggests that in blind cases the \ac{rir} should be jointly estimated with the anechoic speech, which is the paradigm followed by our method BUDDy.

\definecolor{cb1}{HTML}{D81B60}
\definecolor{cb2}{HTML}{1E88E5}
\definecolor{cb3}{HTML}{D29E02}
\definecolor{cb4}{HTML}{004D40}

\newcommand{\xs}{0.06\textwidth}
\newcommand{\ys}{0.3\textwidth}
\newcommand{\w}{0.3\textwidth}
\newcommand{\h}{0.25\textwidth}
\newcommand{\lxs}{-0.14\textwidth}
\newcommand{\lys}{-0.03\textwidth}

\newcommand{\linew}{0.4mm}

\pgfplotsset{
    every axis plot/.append style={line width=2pt},
    discard if not/.style 2 args={
        x filter/.append code={
            \edef\tempa{\thisrow{#1}}
            \edef\tempb{#2}
            \ifx\tempa\tempb
            \else
                \def\pgfmathresult{inf}
            \fi
        }
    }
}

\begin{figure}[t!]
    
\resizebox{\columnwidth}{!}{%
\begin{tikzpicture}
  \begin{axis}[
    xlabel={RIR Error SNR [dB]},
    ylabel={PESQ},
    legend style={at={(0.5,-0.2)}, anchor=north, legend columns=3},
    symbolic x coords={0, 10, 20, 30, 100},
    xtick=data,
    ymin=0.8, ymax=4.1,
    name=pesq,
    width=\w,
    height=\h,
  ]

    \addplot[
        cb1, 
        only marks,
        mark=*,
        discard if not={rir_error_type}{all},
        discard if not={sampler}{kodrasi}
    ] table[x=rir_error_snr, y=pesq, col sep=comma]{data.csv};
    \addplot[
        cb1, 
        line width=\linew, 
        forget plot, %
        discard if not={rir_error_type}{all},
        discard if not={sampler}{kodrasi}
    ] table[x=rir_error_snr, y=pesq, col sep=comma]{data.csv};

    \addplot[
        cb2, 
        only marks,
        mark=*,
        mark options={fill=white}, %
        discard if not={rir_error_type}{all},
        discard if not={sampler}{dps}
    ] table[x=rir_error_snr, y=pesq, col sep=comma]{data.csv};
    \addplot[
        cb2, 
        line width=\linew, 
        loosely dotted, %
        forget plot, %
        discard if not={rir_error_type}{all},
        discard if not={sampler}{dps}
    ] table[x=rir_error_snr, y=pesq, col sep=comma]{data.csv};

  \end{axis}

    \begin{axis}[
    xlabel={RIR Error SNR [dB]},
    ylabel={DNSMOS},
    legend style={at={(0.5,-0.2)}, anchor=north, legend columns=3, xshift=\lxs, yshift=\lys},
    symbolic x coords={0, 10, 20, 30, 100},
    xtick=data,
    ymin=1.5, ymax=4.1,
    name={dnsmos},
    at={(pesq.south east)},
     xshift=\xs,
    width=\w,
    height=\h,
  ]

    \addplot[
        cb1, 
        only marks,
        mark=*,
        discard if not={rir_error_type}{all},
        discard if not={sampler}{kodrasi}
    ] table[x=rir_error_snr, y=dnsmos, col sep=comma]{data.csv};
    \addplot[
        cb1, 
        line width=\linew, 
        forget plot, %
        discard if not={rir_error_type}{all},
        discard if not={sampler}{kodrasi}
    ] table[x=rir_error_snr, y=dnsmos, col sep=comma]{data.csv};

    \addplot[
        cb2, 
        only marks,
        mark=*,
        mark options={fill=white}, %
        discard if not={rir_error_type}{all},
        discard if not={sampler}{dps}
    ] table[x=rir_error_snr, y=dnsmos, col sep=comma]{data.csv};
    \addplot[
        cb2, 
        line width=\linew, 
        loosely dotted, %
        forget plot, %
        discard if not={rir_error_type}{all},
        discard if not={sampler}{dps}
    ] table[x=rir_error_snr, y=dnsmos, col sep=comma]{data.csv};

    \legend{RIF+Post \cite{Kodrasi2014}\; , InfDerevDPS \cite{lemercier2023derevdps}}

  \end{axis}
  
\end{tikzpicture}
}

\caption{\protect\centering\textit{Robustness of informed dereverberation approachess with respect to normally distributed errors in the RIR.
}}
\label{fig:robustness}
\end{figure}

\begin{table}[t!]
    \centering
    \caption{\centering\textit{Dereverberation results on matched reverberant VCTK dataset.
    We indicate for each method in the table if it operates in a blind scenario.
    }}
    \resizebox{\columnwidth}{!}{%
    \begin{tabular}{l|c|cc}
    
Method & Blind & DNS-MOS & PESQ
\\

\midrule
\midrule
Reverberant & -
& 3.14 $\pm$ 0.52
& 1.61 $\pm$ 0.37
\\
\midrule

RIF+Post \cite{Kodrasi2014} & \xmark
& 3.41 $\pm$ 0.47
& 2.66 $\pm$ 0.40
\\

InfDerevDPS \cite{lemercier2023derevdps} & \xmark
& 3.91 $\pm$ 0.33
& 3.95 $\pm$ 0.42
 \\
 
\midrule \midrule

FiNS/RIF+Post \cite{steinmetz2021filtered, Kodrasi2014} & \cmark
& 2.18 $\pm$ 0.38
& 1.33 $\pm$ 0.19
\\

FiNS/InfDerevDPS\cite{steinmetz2021filtered, lemercier2023derevdps} & \cmark
& 2.19 $\pm$ 0.43
& 1.32 $\pm$ 0.18
\\

\midrule

BUDDy (ours) \cite{moliner2024buddy} & \cmark
& \textbf{3.76} $\pm$ \textbf{0.41}
& \textbf{2.30} $\pm$ \textbf{0.53}
\\

\midrule
    \bottomrule
    \end{tabular}
    }
    \label{tab:inf_vs_blind_results}
\end{table}

\subsection{Room Impulse Response Estimation} \label{sec:rir}
\begin{figure*}[t]
    \centering
    \includegraphics{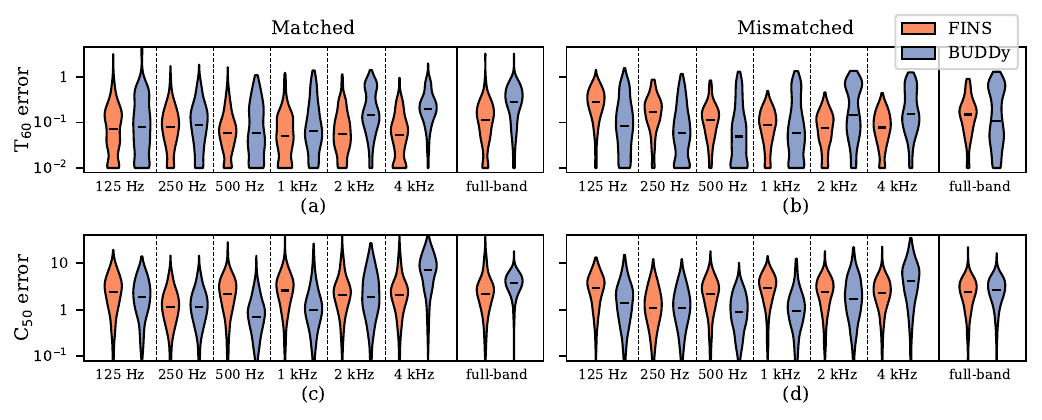}
    \caption{
    \protect\centering\textit{RIR estimation metrics for each octave and full-band on the reverberant VCTK dataset. The violin plots show the distribution and the median. Lower is better. 
    FiNS \cite{steinmetz2021filtered} is trained in a supervision fashion whereas BUDDy is unsupervised.
    }}
    \label{fig:metrics}
\end{figure*}

BUDDy is not only designed as a dereverberation algorithm but also functions as a blind unsupervised RIR estimator. We evaluate its performance for RIR estimation using the same speech model and data we employed for speech dereverberation in Section \ref{sec:vctk}.

\vspace{0.5em}\subsubsection{Baseline}
\label{sec:rir:baseline}

We benchmark BUDDy against FiNS \cite{steinmetz2021filtered}, a DNN-based approach trained to estimate time-domain RIRs directly from reverberant speech.
FiNS comprises a 1D-convolutional encoder and a two-component decoder. 
The first decoder component models the late tail of the RIR by passing noise signals through a trainable filterbank containing several FIR filters.
The second decoder component directly estimates the direct path and early reflections in the time-domain.
In contrast to BUDDy, FiNS relies on supervised learning, thus requiring a paired dataset of reverberant speech and RIRs.
We use an unofficial re-implementation\footnote{https://github.com/kyungyunlee/fins} and train the model on our VCTK-based reverberant speech dataset.

\vspace{0.5em}\subsubsection{Evaluation metrics}
\label{sec:rir:metrics}

Due to the highly ill-posed nature of the blind RIR estimation problem and the statistical nature of late reflections \cite{Habets2010},
we refrain from using element-wise distances, such as error-to-signal ratios, to evaluate the performance of RIR estimators.
Instead, it is arguably more important to preserve the acoustic and perceptual properties of the reference RIR \cite{dalsanto2024similarity}. 
On the other hand, single metrics such as the full-band T$_{60}$ reverberation time or clarity index C$_{50}$ do not account for frequency-specific estimation errors.
We therefore incorporate both full-band and subband reverberation time T$_{60}$ and clarity index C$_{50}$, with subbands spanning octaves. 
This enables to keep a high-level representation of the acoustical properties while allowing enough granularity on the spectral attributes of the RIR.

The reverberation time T$_{60}$ is defined for a diffuse sound field as the time it takes for its energy decay curve (EDC) to decay by $60$\,dB \cite{Naylor2011}.
In order to avoid the effects of the noise floor, 
we calculate T$_{60}$ as twice the time required for the EDC to decrease from $-5$ dB to $-35$ dB relative to the initial level, thereby eliminating the influence of the direct path. 
This measure is computed in each octave band separately.
The octave clarity index C$_{50}$ is the ratio (in dB) between the energy in the first $50$ ms and the energy in the remaining of the RIR, calculated in the corresponding octave band \cite{Naylor2011}. 
Consequently, we compute the absolute error between the T$_{60}$ and C$_{50}$ values calculated for each octave from the estimated RIR and those from the ground truth RIR.

\vspace{0.5em}\subsubsection{Results}
\label{sec:rir:results}

The results for both matched and mismatched test sets are plotted in Fig. \ref{fig:metrics}.  
In the matched condition, FiNS and BUDDy achieve similar T$_{60}$ error rates at low- and mid-range frequency bands, while BUDDy's performance decreases at high frequencies (Fig. \ref{fig:metrics}a). 
Our intuition is that the lower RIR estimation abilities of BUDDy at high frequencies can be related to the tendency of diffusion models to generate high-frequency components in the later stages of the reverse diffusion process \cite{Yang2023DPMMadeSlim}.
Consequently, there is less information available for optimizing the RIR parameters in this range when beginning sampling, negatively affecting parameter convergence.
A similar trend is observed for the C$_{50}$ error in Fig. \ref{fig:metrics}c. Furthermore, BUDDy generally achieves lower C$_{50}$ error than FiNS in the mid-frequency range, where most of the speech content lies.

In the mismatched setting, FiNS struggles to generalize because of its supervised training setup.
As a result, BUDDy outperforms FiNS in both T$_{60}$ and C$_{50}$ error at low and mid-frequency bands (Figs. \ref{fig:metrics}b and \ref{fig:metrics}d). 
At higher frequencies, BUDDy's T$_{60}$ estimation performance still remains slightly inferior to FiNS, though the gap is noticeably smaller than in the matched setting. Regarding C$_{50}$, BUDDy outperforms FiNS in all frequency bands except for the highest 4-kHz band. 
This increased relative performance of BUDDy compared to FiNS highlights the benefits of leveraging unsupervised training for RIR estimation in variable acoustic conditions.

\section{Conclusion}
In this paper, we presented an unsupervised method that simultaneously performs blind dereverberation and \ac{rir} estimation using diffusion models. 
Our results highlight the importance of joint speech and RIR estimation in contrast to plugging estimated RIRs into informed dereverberation methods. 
The proposed method, BUDDy, yields state-of-the-art performance among unsupervised approaches for blind speech and singing voice dereverberation, outperforming both traditional and DNN-based methods. 
Unlike blind supervised methods, which often struggle with generalization to unseen acoustic conditions, our unsupervised approach naturally overcomes this limitation due to its ability to adapt the reverberation operator to a broad range of RIRs.
This holds as well for RIR estimation, as we show that the RIR estimation performance of BUDDy surpasses that of a state-of-the-art supervised DNN-based technique in mismatched acoustic conditions while being on par in a matched setting.

\label{sec:conclusion}

\section*{Acknowledgments}

We would like to thank 
Koichi Saito, Fumikuri Yohena, and Kohei Yatabe for providing us with code and guidance through their methods.
Many thanks to Julius Richter and Till Svajda for their advice on retraining SGMSE+ for singing voice dereverberation.

\bibliographystyle{IEEEbib}
\bibliography{refs23}

\appendix

\subsection{Minimum Phase Constraint} \label{sec:appendix_minphase}

The minimum-phase constraint in Section \ref{sec:operator} takes the time-domain RIR $\mathbf{h}$ and computes the minimum-delay phase $\Theta$ as
\begin{equation}
    \Theta = -  \Im \left[\mathcal{H} \left( \log | \mathbf{\mathcal{F}(\mathbf{h}) | } \right) \right],
\end{equation}
where $\mathcal{F}$ is the Fourier transform and $\mathcal{H}$ the Hilbert transform:
\begin{equation}
    \mathcal{H}(\mathbf{x}) \overset{\Delta}{=} \mathcal{F}^{-1}(-j \cdot \mathrm{sign}(\omega)\mathcal{F}(\mathbf{x})).
\end{equation}
The minimum-delay corrected time-domain RIR is then obtained by replacing the original phase with the obtained minimum-delay phase:
\begin{equation}
    \mathbf{h}_\text{min}= \mathcal{F}^{-1}(|\mathcal{F}(\mathbf{h})| e^{j \Theta }).
\end{equation}
All the operations involved in this method are differentiable, which allows backpropagation throughout the process.

\subsection{Noise Regularization} \label{sec:appendix_noisereg}

Section \ref{sec:rir-optimization} introduces a noise regularization term, which we can simplify ignoring scaling factors as
\begin{equation}\label{eq:noise_reg_appendix}
\mathcal{R}(\psi)=
\rVert \
 S_\mathrm{comp}(
\mathbf{h}_\psi)
-
S_\text{comp}(
\mathrm{sg} \left[ \mathbf{h}_\psi \right] +\nu(\tau) \mathbf{v})\
\lVert_2^2 .
\end{equation}

\noindent
The gradient computed during optimization is obtained as
\begin{align}
    \frac{\partial \mathcal{R}(\psi) }{\partial \psi} &=
    2
    \left(
    S_\mathrm{comp}(\mathbf{h}_\psi)- S_\mathrm{comp}(\mathrm{sg} \left[\mathbf{h}_{\psi}\right]+ \nu(\tau)\mathbf{v}) 
    \right) \nonumber \\
    & \quad \times
    \frac{\partial S_\mathrm{comp}}{\partial \mathbf{h}_\psi}
    \times
    \left( \frac{\partial \mathbf{h}_\psi}{ \partial \psi}-\underbrace{ \frac{\partial \, \mathrm{sg} \left[\mathbf{h}_{\psi}\right]}{ \partial \psi} }_{0}\right) \nonumber \\
    & \approx
    -2 \sigma^\prime \mathbf{v} \left[ \frac{\partial S_\mathrm{comp}}{\partial \mathbf{h}_\psi} \right]^2 
    \frac{\partial \mathbf{h}_\psi}{ \partial \psi}, \nonumber
\end{align}
where we have ignored second- and higher-order Taylor expansion terms of $S_\mathrm{comp}$ for simplicity.
We observe that the resulting gradient for $\mathcal{R}(\psi)$ is proportional to the noise vector $\mathbf{v}$ and to the gradient of the estimated RIR $\mathbf{h}(\psi)$ with respect to the parameters $\psi$. Therefore, adding $\mathcal{R}(\psi)$ in the optimization has the result of adding multiplicative noise to the operator gradients (with respect to $\psi$) which emerge from the optimization of the reconstruction loss $\mathcal{C}(\reverb, \mathcal{A}_\psi(\clean))$. Empirically, this has the effect of smoothing out the optimization of the RIR operator parameters $\psi$ and avoiding degenerate solutions, provided that the dedicated noise schedule $\nu(\tau)$ is reasonably chosen.

\subsection{Experimental Details} 
\label{sec:appendix_exp}

\subsubsection{Speech Dereverberation}
\label{sec:appendix_exp_vctk}

\paragraph{Architecture and training hyperparameters} 
\label{sec:appendix_exp_vctk:hp}

We train the unconditional score model $\mathbf{s}_\theta$ for our method BUDDy with anechoic data only, using segments of 4\,s randomly extracted from the utterances in VCTK.
Same as in \cite{moliner2024buddy, lemercier2023derevdps}, we implement the unconditional score network architecture with NCSN++M\cite{Lemercier2022analysing,Lemercier2022storm},
a lighter variant of the NCSN++ \cite{song2021sde} with 27.8M parameters.
Similar to \cite{moliner_solving_2022}, we wrap up the network with a time-frequency transform, in this case the STFT,
such that the NCSN++M forward pass is effectively performed in the complex STFT domain
using a real and imaginary parts representation.
For all methods, STFTs are computed using a Hann window of 32\,ms and a hop size of 8\,ms. 
The complex prediction at every state can be converted to time-domain by inverting the STFT.
We adopt Adam \cite{kingma2015adam} as the optimizer to train the unconditional score model, 
with a learning rate of $10^{-4}$ and an
effective batch size of 16 for 200k iterations. 
We track an exponential moving average of the DNN weights with a decay of 0.999 to be used for sampling 
as in \cite{richter2023speech}.

\noindent
\paragraph{Reverberation operator}
\label{sec:appendix_exp_vctk:hp-op}

The STFT parameters are the same as those used in the unconditional score model, i.e. we use a Hann window of 32 ms and a hop size of 8 ms. 
For subband filtering we further employ 50\% zero-padding to avoid frequency aliasing artifacts. 
Given our sampling rate of $f_s=16$ kHz, this results in $K=513$ unique frequency bins. We set the number of STFT frames of our operator to $N_\mathbf{h} = 100$ (800\,ms).
We subsample the frequency scale in $B=26$ bands, with a $125$-Hz spacing between $0$ and $1$\,kHz, a $250$-Hz spacing between $1$ and $2$\,kHz, and a $500$-Hz spacing between $3$ and $8$\,kHz.

We optimize the \ac{rir} parameters $\psi$ using Adam, with a learning rate of 0.1, and the momentum parameters are set to $\beta_1=0.9$, and $\beta_2=0.99$. 
We employ $N_\text{its.}=10$ optimization iterations per diffusion step.
We further constrain the weights $w_b$ between 0 and 40\;dB, and the decays $\alpha_b$ between 0.5 and 28. This avoids the optimization from approaching degenerate solutions, especially at the early stages of sampling.

\noindent
\paragraph{Forward and reverse diffusion}
\label{sec:appendix_exp_vctk:hp-diff}

As mentioned in Section \ref{sec:posterior-speech} we obtain our initial estimate $\mathbf{x}_\text{init}$ through WPE dereverberation. 
Consequently, we choose $T=0.5$ such that the initial noise in $\mathbf{x}_T \sim \mathcal{N}(\mathbf{x}_\text{init}, \sigma^2(T) \mathbf{I})$ effectively masks potential artifacts stemming from WPE, while still retaining the general structure in $\mathbf{x}_\text{init}$ that may guide the process.
We set the minimal diffusion time to $T_\mathrm{min}=10^{-4}$ and adopt the same reverse discretization scheme as Karras et al. \cite{karras2022elucidating}:
\begin{equation}
\resizebox{0.89\hsize}{!}{
    $\forall i < N,  \, \tau_i = \sigma_i = \left(T^{1/\rho} + \frac{i}{N-1} ( T_\mathrm{min}^{1/\rho} - T^{1/\rho} )\right)^\rho,$
    }
\end{equation}
with warping $\rho=10$ and $N=200$ steps. We use the second-order Euler-Heun stochastic sampler in \cite{karras2022elucidating} with $S_\mathrm{churn}=50$.
In the noise regularization term depicted in \eqref{eq:noise_reg}, 
the annealing schedule $\nu(\tau)$ follows the same discretization as $\sigma(\tau)$, but we restrict its values between $\sigma^\prime_\text{min} = 5\times10^{-4}$ and $\sigma^\prime_\text{max}=10^{-2}$.
The scaling factor used for the variance estimate $\eta(\tau)$ in \eqref{eq:variance} is fixed to $\tilde{\eta}=0.5$.

\noindent
\paragraph{Blind Baselines}
\label{sec:appendix_exp_vctk:baselines}

For WPE \cite{Nakatani2008a}, we take 5 iterations, a filter length of 50 STFT frames (400\,ms) and a delay of 2 STFT frames (16\,ms).
We set the hyperparameters of the method by Yohena and Yatabe \cite{yohena2024single} to $M=50$ and $\rho=400$ after conducting a parameter search. 
Using code gently provided by the authors, we retrain Saito et al. \cite{saito2023unsupervised} and GibbsDDRM \cite{murata_gibbsddrm_2023} using the same data as for BUDDy, i.e. the anechoic VCTK dataset. We use the same inference parameters which can be found in \cite{saito2023unsupervised, murata_gibbsddrm_2023} although we tried to improve the results by doing a hyperparameter search as suggested by the authors.
We re-train RVAE-EM in unsupervised mode on our anechoic VCTK dataset using publicly available code and use the original inference parameters reported by the authors \cite{Wanf2024rvaeem}.

\noindent
\paragraph{Informed Baselines}
\label{sec:appendix_exp_vctk:informed_baselines}

The informed dereverberation method InfDerevDPS is described in Section \ref{sec:dps}.
The scaling factor used for the variance estimate $\eta(\tau)$ in \eqref{eq:variance} is increased to $\tilde{\eta}=2.75$ compared to the blind case, as more confidence can be allocated to the likelihood model.
The informed baseline RIF+Post \cite{Kodrasi2014} uses a regularized inverse filter with a regularization factor of $\delta = 0.01$. The utterance is then post-processed using a Wiener filter with an \textit{a priori} SNR obtained via \cite{Breithaupt2008novel} to remove pre-echoes.

\noindent
\paragraph{Listening experiment}
\label{sec:appendix_exp_vctk:listening}

We conducted a listening experiment based on the MUSHRA recommendation \cite{ITURmushra} using the \texttt{webMUSHRA}\footnote{https://github.com/audiolabs/webMUSHRA} interface. 
The test comprised 12 pages, featuring 6 reverberant speech utterances from each of the \emph{matched} and \emph{mismatched} datasets.
The test was conducted in isolated conditions within listening booths at the Aalto Acoustics Lab. In total, 10 volunteers participated in the experiment. 
All utterances were loudness-normalized to -23dB LUFS.
The participants were allowed to modify the volume of headphones during the training stage (first page, not included in the results).
The ground-truth anechoic speech served as the reference, which was also hidden among the other conditions (WPE, RVAE-EM, PSE, SGMSE, BUDDy), while the original reverberant speech signal was used as the low anchor, expected to receive a score of 0.
Participants were advised to focus particularly on dereverberation performance and to use the full rating scale, i.e., rate the reference as 100 and the reverberant anchor as 0.
We obtained consent directly from the participants through a written form. As the study did not present any risk for the subjects, no review board was required for the approval of this experiment.

\subsubsection{Singing Voice Dereverberation} 
\label{sec:appendix_exp_nhss}

We use the same NCSN++-based architecture as in the speech dereverberation experiments, and modify the \ac{stft} parameters to account for the new sampling frequency. Specifically, we employ a $1534$-point window and hop size of $384$.
The unconditional score model is optimized using Adam with same parameters as for the VCTK dataset, but we reduce the batch size to $4$ and use 6-s anechoic audio segments. 
We use $B=39$ bands for the subband decomposition in the reverberation operator for BUDDy, extending the bands used in Appendix \ref{sec:appendix_exp_vctk:hp-op} above $8$\,kHz with a $1$\,kHz spacing.
Because of implementation reasons, the SGMSE+ and PSE supervised baselines are trained using $48$-kHz-resampled data, and utterances are downsampled to $44.1$\,kHz after processing.

\begin{IEEEbiography}[{\includegraphics[width=1in,height=1.25in,clip,keepaspectratio]{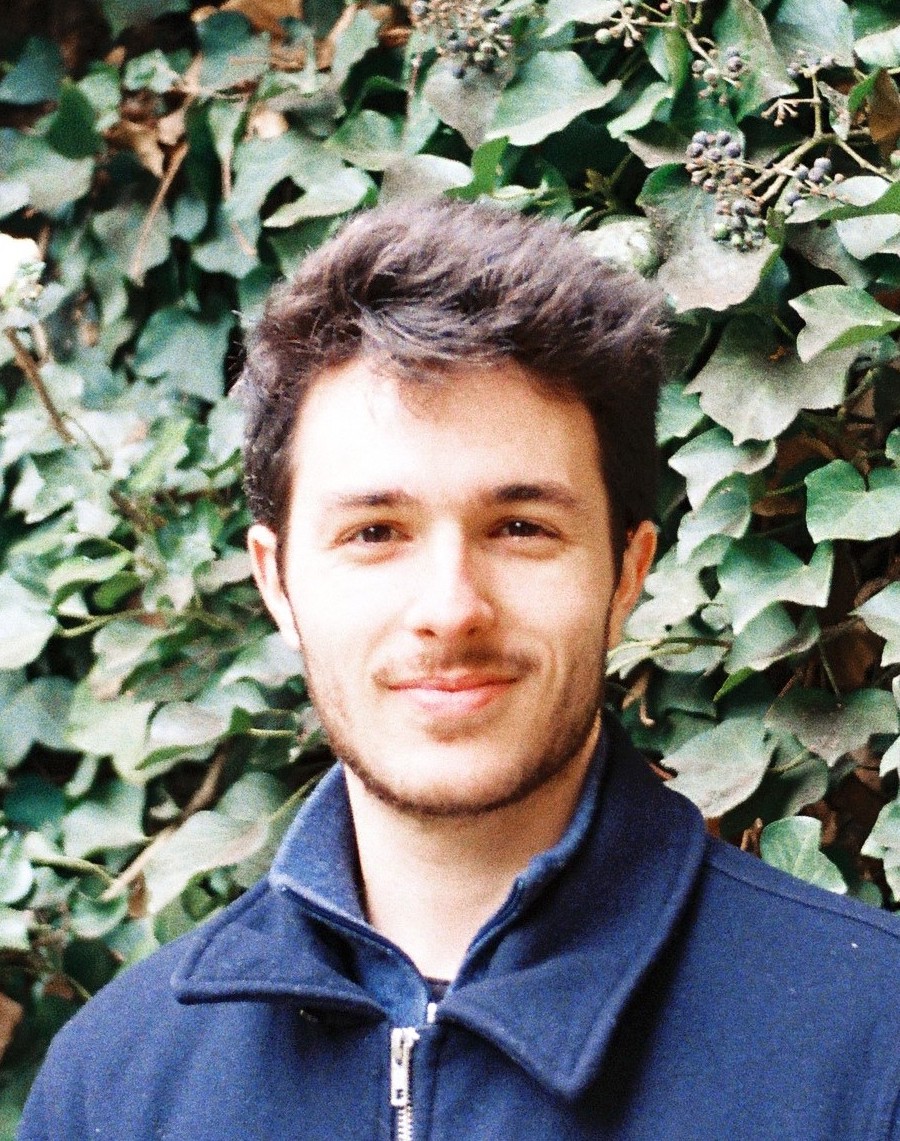}}]{Jean-Marie Lemercier}
received an M.Eng in Electrical Engineering in 2019 from Ecole Polytechnique, Paris, France. In 2020, he received a M.Sc. in Communications and Signal Processing from Imperial College London, London, UK. He is currently a PhD student in the Signal Processing group at Universität Hamburg under the supervision of Prof. Dr.-Ing. Timo Gerkmann. His research interests span machine learning-based speech enhancement and dereverberation for hearing devices applications. Recent works also include the design and analysis of diffusion-based generative models for various speech restoration tasks. He is a Student Member of IEEE. He is a recipient of the VDE ITG 2024 award.
\end{IEEEbiography}

\begin{IEEEbiography}[{\includegraphics[width=1in,height=1.25in,clip,keepaspectratio]{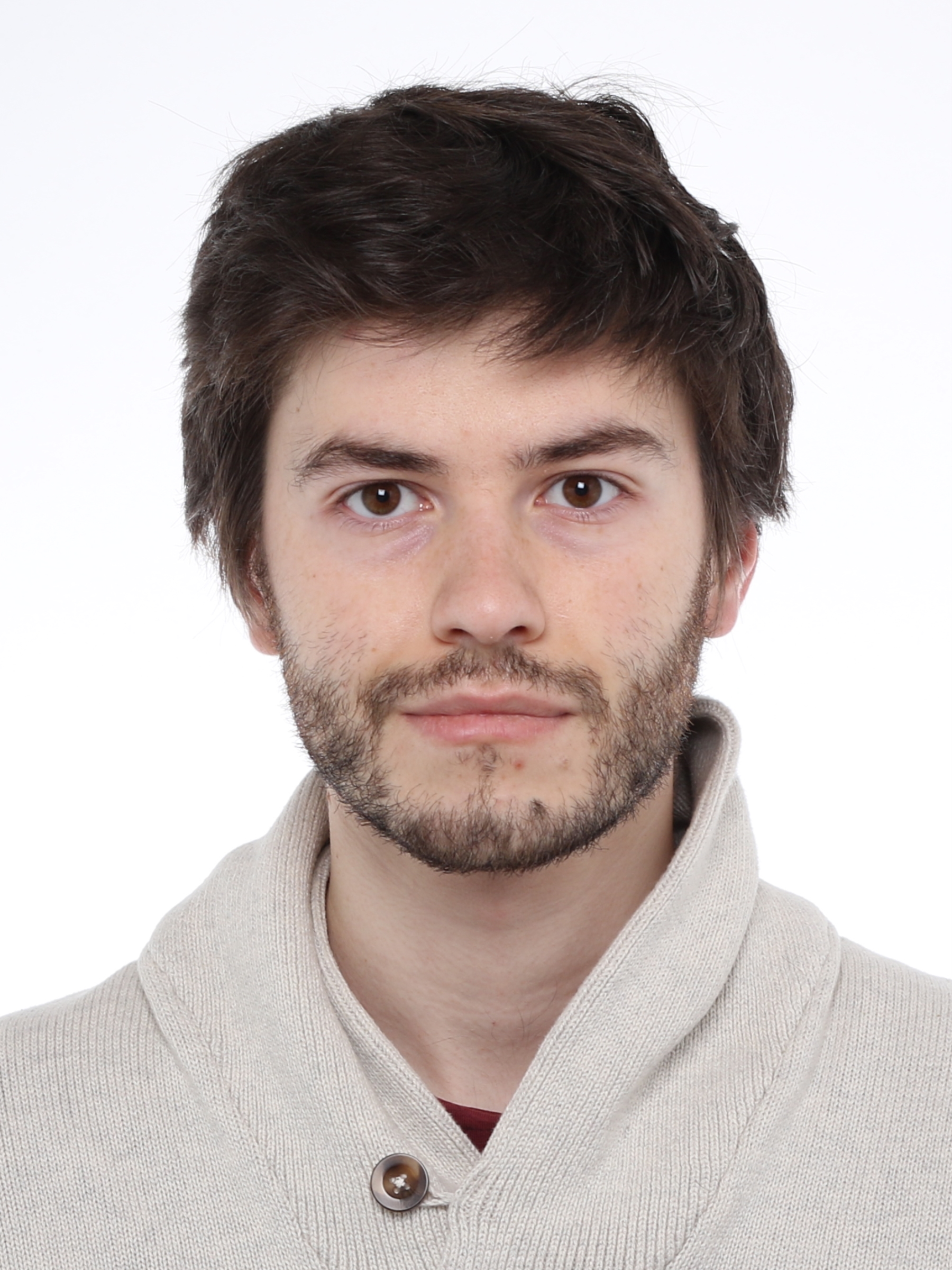}}]{Eloi Moliner}
 received his B.Sc. degree in Telecommunications Technologies and Services Engineering from the Polytechnic University of Catalonia, Spain, in 2018 and his M.Sc. degree in Telecommunications Engineering from the same university in 2021. He is currently a doctoral candidate at the Acoustics Lab of Aalto University in Espoo, Finland. His research interests include digital audio restoration and audio applications of machine learning. He is the recipient of the Best Student Paper Award of the 2023 IEEE ICASSP conference.

\end{IEEEbiography}

\begin{IEEEbiography}[{\includegraphics[width=1in,height=1.25in,clip,keepaspectratio]{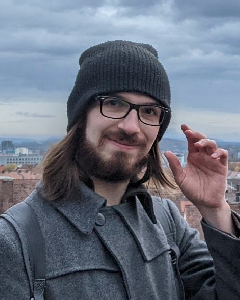}}]{Simon Welker}
received a B.Sc. in Computing in Science (2019) and an M.Sc. in Bioinformatics (2021) from Universität Hamburg, Germany. He is currently a PhD student under the supervision of Prof. Dr.-Ing. Timo Gerkmann (Signal Processing, Universität Hamburg) and Prof. Dr. Dr. Henry Chapman (Center for Free-Electron Laser Science, DESY, Hamburg), researching machine learning techniques for solving inverse problems that arise in speech processing and X-ray imaging. 

\end{IEEEbiography}

\begin{IEEEbiography}[{\includegraphics[width=1in,height=1.25in,clip,keepaspectratio]{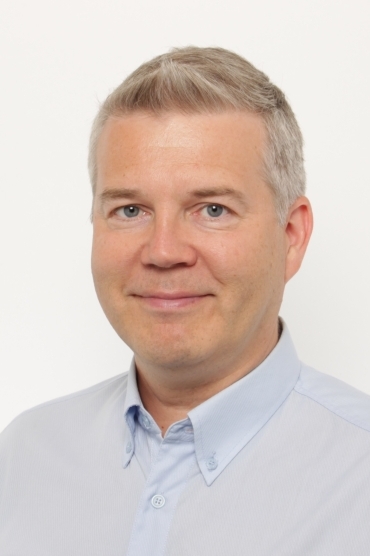}}]{Vesa V\"alim\"aki}
(Fellow, IEEE) received his D.Sc. degree in electrical engineering from the Helsinki University of Technology in 1995. In 1996, he was a post-doctoral research fellow at the University of Westminster, London, UK. In 2008–2009 he was a visiting scholar at Stanford University. He currently is a Professor of audio signal processing and Vice Dean for Research at Aalto University, Espoo, Finland. His research interests include the application of machine learning and signal processing to audio technology. He is a Fellow of the Audio Engineering Society and the Asia-Pacific Artificial Intelligence Association. He is the Editor-in-Chief of the \emph{Journal of the Audio Engineering Society}.

\end{IEEEbiography}

\begin{IEEEbiography}[{\includegraphics[width=1in,height=1.25in,clip,keepaspectratio]{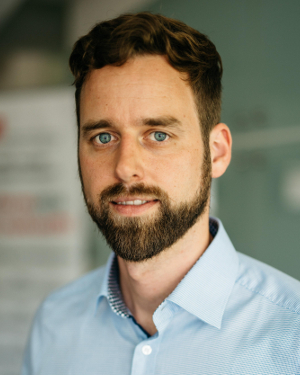}}]{Timo Gerkmann}
is a Professor for Signal Processing with the Universität Hamburg, Hamburg, Germany. He has held positions with Technicolor Research \& Innovation, University of Oldenburg, Oldenburg, Germany, KTH Royal Institute of Technology, Stockholm, Sweden, Ruhr-Universität Bochum, Bochum, Germany, and Siemens Corporate Research, Princeton, NJ, USA. His research interests include statistical signal processing and machine learning for speech and audio applied to communication devices, hearing instruments, audio-visual media, and human-machine interfaces. He was the recipient of the VDE ITG award 2022. He served in the IEEE Signal Processing Society Technical Committee on Audio and Acoustic Signal Processing and is currently a Senior Area Editor of the IEEE/ACM Transactions on Audio, Speech and Language Processing.

\end{IEEEbiography}

\newpage

\end{document}